\documentstyle[multicol,eqsecnum,prl,aps,epsf]{revtex}
\input epsf
\newcommand{\be}{\begin{equation}}
\newcommand{\ee}{\end{equation}}
\newcommand{\bea}{\begin{eqnarray}}
\newcommand{\eea}{\end{eqnarray}}

\newcommand{\w}{\omega}
\newcommand{\vk}{ {\bf k} }
\newcommand{\e}{\epsilon}
\begin{document}
\draft
\title{Critical scaling of the a.c.\ conductivity \\ for a superconductor
above $T_{c}$}
\author{Robert A. Wickham\cite{BYLINE1} and Alan T. Dorsey\cite{BYLINE2}}
\address{Department of Physics, University of Florida \\
 P.O. Box 118440,
Gainesville, Florida  32611-8440}
\date{\today}
\maketitle
%
%

\begin{abstract}

We consider the effects of critical superconducting fluctuations 
on  the scaling of the linear a.c.\ conductivity,  $\sigma(\omega)$, of  
a bulk superconductor slightly above $T_{c}$ in zero applied 
magnetic field. The dynamic renormalization-group method is applied to 
the relaxational time-dependent Ginzburg-Landau model of 
superconductivity, with $\sigma(\omega)$ calculated {\em via} the
Kubo formula to $O(\e^{2})$ in the $\e =  4 - d$ expansion.
The critical dynamics are governed by the relaxational XY-model 
renormalization-group fixed point.
The scaling hypothesis $\sigma  (\omega)  \sim \xi^{2-d+z} 
S(\omega   \xi^{z})$ proposed  by Fisher, Fisher and Huse is  explicitly
verified, with the dynamic exponent $z \approx 2.015$, the value
 expected for the $d=3$ relaxational XY-model. The 
universal scaling function $S(y)$  is computed and shown to 
deviate only slightly from its Gaussian form, calculated earlier.
The present theory is compared with experimental measurements of the 
a.c.\ conductivity of YBCO near $T_{c}$, and the implications of this
theory for such experiments is discussed.   

\end{abstract}
\pacs{PACS number(s): 74.25.Fy, 74.40.+k, 64.60.Ht}
\begin{multicols}{2}
%
%

\section{Introduction}

The discovery of high-temperature superconductors has, for the first time, 
made it possible to experimentally probe the critical region of 
the zero-field normal-superconducting 
transition since fluctuation effects 
in these materials are enhanced by the short coherence length and 
the  high  transition temperature $T_{c}$. 
It is natural then to 
ask: If scaling and universality exist in the critical region, 
to which universality class does the transition belong?
From observations of the effects of critical superconducting  
fluctuations on thermodynamic 
properties, such as the penetration depth \cite{KAMAL94,ANLAGE96}, 
magnetic susceptibility \cite{SALAMON93,POMAR93,LIANG96},
specific heat \cite{SALAMON93,OVEREND94}
and thermal expansivity \cite{PASLER98} 
a consensus is emerging  that the zero-field normal-superconducting 
transition is in the {\em static} universality 
class of the three-dimensional, complex order-parameter (3D XY) model. 
In contrast, the
 effect of critical fluctuations on transport properties, such as 
the conductivity, depends on the nature of the {\em dynamics} near $T_{c}$
and is much less explored. 

In general, conductivity measurements on high-$T_{c}$ 
superconductors show an enhanced response above $T_{c}$ due to 
the presence of superconducting fluctuations. Outside the critical region 
this enhancement can be explained in terms of the Aslamazov-Larkin 
\cite{ASLAMAZOV68} theory of non-interacting, Gaussian fluctuations,
and its extensions \cite{SCHMIDT68,DORSEY91}. In these 
theories the dynamic exponent $z$ associated with the growth 
of the characteristic order-parameter time-scale near $T_{c}$
appears in the conductivity and takes the value $z=2$. By examining the
deviation of $z$ from $2$ inside the critical region through 
linear d.c. \cite{SALAMON93,POMAR93,HOWSON95,HOLM95,MOLONI97a,KIM97},
non-linear d.c. \cite{DEKKER91,ROBERTS95,MOLONI97b}
and linear a.c.  \cite{BOOTH96} conductivity measurements,
the dynamic universality class can, in principle, be determined. 
Currently, however, there is much variation in the  measured values for 
$z$ and the dynamic universality class of the zero-field 
normal-superconducting transition remains uncertain. Unlike d.c.\ 
measurements,
measurements of the a.c.\ conductivity \cite{BOOTH96}  can test the
scaling of the conductivity, $\sigma (\w)$, over a wide range of frequencies,
 $\w$, thereby providing a stringent test of theory.
In the experiments of Ref. \cite{BOOTH96} the a.c.\ conductivity 
exhibits a scaling collapse which deviates slightly from the Gaussian 
theory. However, 
the Gaussian theory is known 
to break down in the critical region. Thus, 
to sharpen the comparison between experiment and theory, 
we go beyond the Gaussian 
description of fluctuations in this paper and 
calculate the scaling behaviour of the a.c.\ conductivity in 
the critical region of strong, interacting fluctuations.

Fisher, Fisher and Huse (FFH) \cite{FFH} have argued
that near a second-order phase transition, if dynamic scaling holds,
the a.c.\ fluctuation conductivity should scale as
%
%
\be
\label{EQ:SCALCON}
\sigma(\omega) \sim \xi^{2-d+z} S( \omega \xi^{z} ),
\ee
where the correlation length 
for  fluctuations  in  the superconducting order-parameter at 
temperature $T$ is $\xi \sim |T-T_{c}|^{-\nu}$ with the 
static exponent $\nu$,
$d$ is the spatial dimensionality, 
$z$ is the  dynamic exponent and
$S(y)=S'(y)+ i S''(y)$ 
is a  universal, complex function of the scaled frequency
$y \sim \omega \xi^{z} $, with real and imaginary parts $S'$ and $S''$, 
respectively. 
Outside the critical region, and 
in the d.c.\ limit, Eq.\ (\ref{EQ:SCALCON}) reduces to the 
Aslamazov-Larkin theory.
Since the conductivity is causal, and also finite for 
non-zero frequencies, Eq.\ (\ref{EQ:SCALCON}) leads 
to the power-law behaviour at $T_{c}$
%
%
\be
\label{EQ:CONTC}
\sigma(\omega) \sim (-i \omega)^{-(2-d+z)/z},
\ee
reflecting the absence of a characteristic time-scale 
at criticality.  At $T_{c}$ the phase 
\be
\label{EQ:PHASEDEFN}
\phi(\omega) = 
\tan^{-1} \left(\frac{S''(\omega \xi^{z})}{S'(\omega \xi^{z})} \right)
\ee 
of the conductivity is independent of frequency,
with the value \cite{DORSEY91}
%
%
\be
\label{EQ:PHASE}
\phi = \frac{\pi}{2} \left( \frac{2-d+z}{z} \right).
\ee
Equations (\ref{EQ:CONTC}) and (\ref{EQ:PHASE}) allow one to 
determine the dynamic exponent $z$ independently of the static exponent 
$\nu$ through a measurement of the a.c.\ conductivity at criticality.
However, to go beyond these two results and calculate the entire 
universal scaling function $S(y)$ requires knowledge of the 
renormalization-group fixed-point that determines the universality class
for the dynamics near $T_{c}$.

The time-dependent Ginzburg-Landau (TDGL) model 
of superconductivity provides an 
appropriate framework in which to study dynamic critical behaviour in 
this system \cite{MA76,HH}. 
Since this is the first detailed study of the dynamics in the critical 
region of the superconductor, and given the uncertainty as to 
which dynamic universality class describes the transition,
we consider here only the simplest,
relaxational, dynamics for fluctuations in the superconducting 
order-parameter ---  model A in the 
Hohenberg and Halperin classification 
\cite{HH,SUPERFLUID}. Previous studies of this
model have implemented the
Gaussian approximation, where quartic interactions among fluctuations in the 
Ginzburg-Landau free-energy are neglected \cite{SCHMIDT68,DORSEY91}.
In this approximation,
the conductivity scales as  Eq.\ (\ref{EQ:SCALCON}) with $\nu = 1/2$ and
$z=2$, the exponents for the Gaussian fixed-point, 
and the scaling function $S(\omega \xi^{2})$ 
has been explicitly calculated.
%
%
\begin{figure}
\begin{center}
    \leavevmode
    \epsfxsize=3.25 in
    \epsffile{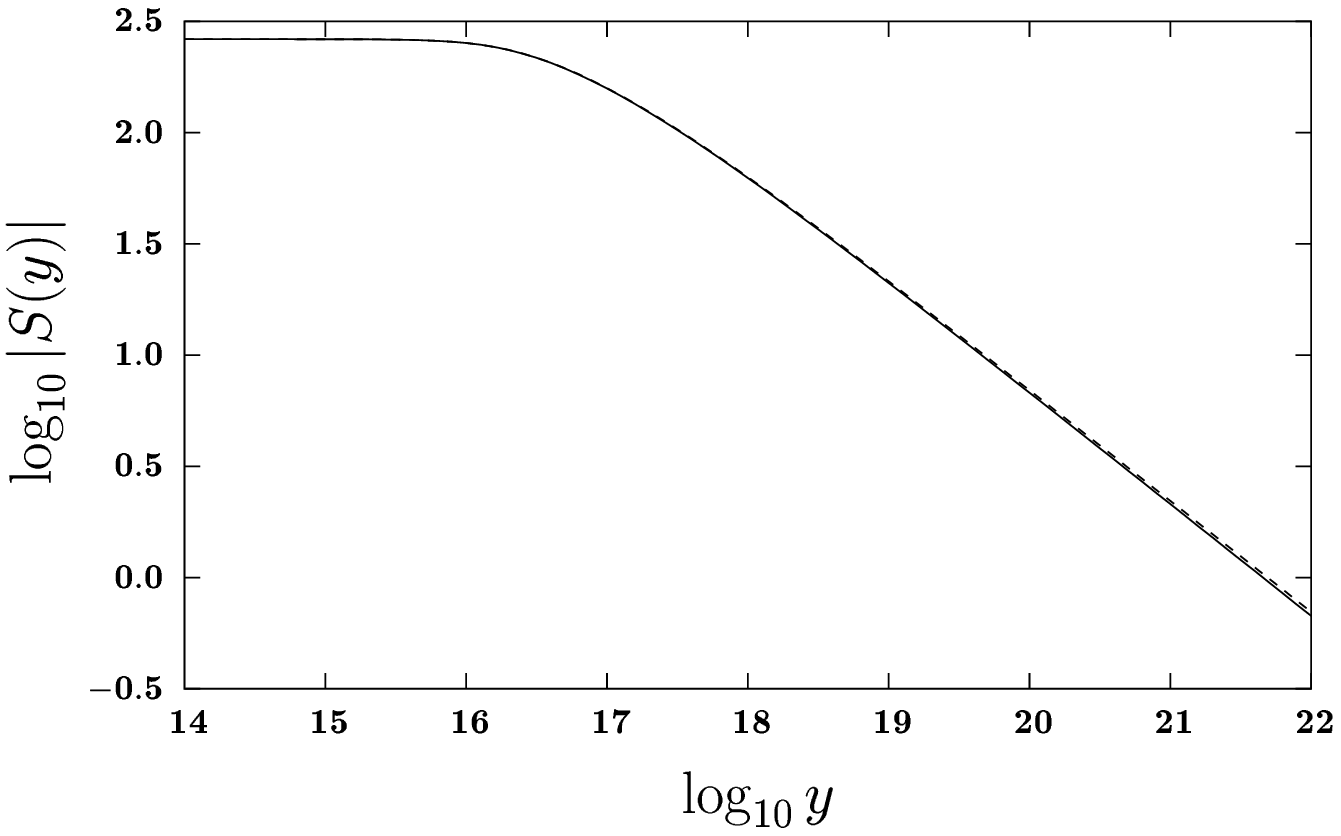}
  \end{center}

\small{FIG. 1. Comparison of the a.c.\
conductivity scaling function $S(y)$, Eq.\ (\ref{EQ:SINTRO}),
for the relaxational 3D XY critical theory (solid curve) with the 
scaling function, Eq.\ (\ref{EQ:CGI}), for the Gaussian theory (dashed curve).
To facilitate later comparison with experiments \protect\cite{BOOTH96},
the magnitude of $S(y)$ is plotted against the scaled 
frequency $y$ on a $\log-\log$ scale.}
\label{FIG:CRITVGAUS}
\end{figure}

In the critical region the Gaussian approximation breaks down since  
the quartic interactions become important, producing the critical
fixed-point for the relaxational XY-model \cite{WILSON72,HHM}.
In the $\e=4-d$ expansion, the exponents for this fixed-point 
are well known \cite{MA76}.
An extrapolation of the $O(\e^{2})$ results to three dimensions
gives a correlation-length exponent of $\nu \approx 2/3$ and a
correlation function exponent of $\eta \approx 0.02$.
For relaxational dynamics the dynamic exponent $z$ is, to $O(\e^{2})$ 
\cite{HHM}:
\be
\label{EQ:ZED}
z = 2 + c \eta
\ee
with
%
%
\be
\label{EQ:C}
c = 6 \ln 4/3 - 1,
\ee
giving $z \approx 2.015$ in three dimensions.

In the critical region, and 
near four dimensions, we verify that the a.c.\
fluctuation conductivity satisfies the FFH 
scaling hypothesis (\ref{EQ:SCALCON}) for the relaxational XY-model
fixed-point.
We compute the complex scaling form $S(y)$ appearing in Eq.\ 
(\ref{EQ:SCALCON}) to $O(\e^{2})$, with the result 
\end{multicols}
\vspace{-.6cm}\noindent\makebox[8.6cm]{\hrulefill}\vspace{-.2cm} \\
\vspace{.15 in}
\be
\label{EQ:SINTRO}
S(y) = \frac{2 z^{2}}{(d-2+z)(d-2)} \frac{1}{y^{2}} 
\left[ 1 - \frac{d-2+z}{z} \mbox{ } iy 
- ( 1 - i y)^{(d-2+z)/z} \right],
\ee
\vspace{.15 in} \\
\vspace{-1.7cm}\noindent\hspace{9.2cm}\makebox[8.6cm]{\hrulefill}\vspace{1.3cm}
\begin{multicols}{2} \hspace{-.2 in}
where $y \sim \w \xi^{z}$ and $z$ is given by  
Eq.\ (\ref{EQ:ZED}) with Eq.\ (\ref{EQ:C}). 
This is the main result of this paper, and is the product of a much more 
involved analysis than that used to determine the exponent $z$.
Sections \ref{SEC:FORMALISM}-\ref{SEC:RENORM} 
provide the details of the calculation.
The result (\ref{EQ:SINTRO}) has the scaling behaviour stated in Eq.\ 
(\ref{EQ:CONTC}). 
Since, to $O(\e^{2})$, $z$ for the critical theory in three 
dimensions is only slightly different than
two, the scaling function $S(y)$ for the critical theory is very close 
to the Gaussian result calculated earlier (see Fig.\ 1). 
In Sec.\ \ref{SEC:COMPARE} we compare the experimental 
a.c.\ conductivity data of Booth {\em et al.} \cite{BOOTH96} to the critical 
theory, extrapolated to three dimensions, 
and comment in Sec.\ \ref{SEC:CONCLUSIONS} on the implications of this work 
for such measurements.
%
%
\section{Formalism}
\label{SEC:FORMALISM}

\subsection{The time-dependent Ginzburg-Landau model
 of superconductivity}

We describe the critical dynamics of a superconductor 
with a complex order-parameter $\psi$ using the 
relaxational time-dependent Ginzburg-Landau model
%
%
\be
\label{EQ:GLMOD}
\frac{\partial \psi} {\partial t} = - \Gamma_{0} 
\frac{\delta F}{\delta \psi^{*}} + \zeta
\ee
with the Ginzburg-Landau free-energy
%
%
\be
\label{EQ:FREE}
F = \int d^{d} r \mbox{ } \left( \mbox{ } 
 |\nabla \psi |^{2} + r_{0} | \psi |^{2}
+ \frac{u_{0}}{2} |\psi|^{4} \right) .
\ee
In Eq.\ (\ref{EQ:GLMOD}) $\Gamma_{0}$ is the bare order-parameter relaxation 
rate. Both $\Gamma_{0}$ and the bare coefficient $u_{0}$, which appears 
in the free-energy (\ref{EQ:FREE}), can be considered temperature-independent
near the transition; however $r_{0} \sim T - T_{c0}$ changes sign at 
the mean-field transition temperature $T_{c 0}$, 
becoming negative for temperatures below $T_{c 0}$. We choose units so that 
$\hbar=k_{B} T_{c}=1$ and $m=1/2$, where $m$ is the mass of a 
Cooper pair. The superconductor is assumed to be isotropic. 
The complex noise field $\zeta$ in Eq.\ (\ref{EQ:GLMOD}) is taken to 
have zero mean and correlations described by
%
%
\be
\label{EQ:NOISE}
\langle \zeta ({\bf r}, t) \zeta^{*}({\bf r}', t') \rangle 
=  2 \Gamma_{0} \mbox{ } \delta({\bf r} -{\bf r}' ) \delta(t- t'),
\ee
where the brackets $\langle \cdots \rangle$ denote an average over 
the noise distribution, assumed to be Gaussian.
The factor $2 \Gamma_{0}$ in Eq.\ (\ref{EQ:NOISE}) follows from the
fluctuation-dissipation theorem and ensures that the system relaxes to 
the proper equilibrium distribution.

We will work in the symmetric phase, $T > T_{c}$, with zero applied 
magnetic field and consider order-parameter fluctuations about a mean
of zero. Fluctuations of the vector potential 
are neglected \cite{MAGFLUCTS}. Since
we will use the Kubo formula to calculate the linear conductivity from
the system in zero electric field, an electric field is
not included
in Eqs.\ (\ref{EQ:GLMOD}-\ref{EQ:FREE}). In the classification 
scheme of Hohenberg and Halperin \cite{HH}, 
Eqs.\ (\ref{EQ:GLMOD}-\ref{EQ:NOISE}) constitute model A dynamics for a
two-component (complex) order-parameter. Thus our model is in the 
dynamic universality class of the relaxational XY-model 
\cite{WILSON72,HHM}.

Since the Ginzburg-Landau theory is coarse-grained, it contains an
ultra-violet (UV) cutoff, $\Lambda$ (corresponding, for example, to the
lattice constant) \cite{MAKITHOM}. This cutoff is manifest in the 
definition of the Fourier transform of the
order-parameter,
%
%
\be
\label{EQ:FT}
\psi ({\bf r},t) = \int_{k \w}^{\Lambda} \psi (\vk, \omega ) \mbox{ }
e^{ i {\bf k} \cdot {\bf r} - i \omega t}.
\ee
For convenience, we employ the short-forms 
%
%
\bea
\int_{k}^{\Lambda} & = & \int^{\Lambda} \frac{d^{d} k}{(2 \pi)^{d} } \\
\int_{\w} & = & \int \frac{d \w}{(2 \pi)}
\eea
for the wavevector and frequency integrals, with the wavevector integral
restricted to $|{\bf k}| < \Lambda$.
The existence of the cutoff will be crucial when we 
 interpret the results of the theory.

The order-parameter correlation function and the response function 
are central in what follows. 
The order-parameter correlation function, $C(\vk,\omega)$,
is defined as 
%
%
\be
\label{EQ:OPCOR}
C(\vk,\omega) \equiv \langle \psi ( \vk,\omega) 
\psi^{*} (\vk,\omega) \rangle.
\ee
By adding a source term, 
%
%
\be
\label{EQ:SOURCE}
F_{h} = - \int d^{d} r \mbox{ } ( h^{*} \psi + h \psi^{*}),
\ee
to the free-energy (\ref{EQ:FREE}) we can define 
the (linear) response function, $G(\vk,\omega)$, as
%
%
\be
\label{EQ:RES}
G({\bf k}, \omega) \equiv \left. \frac{\delta \langle 
\psi ({\bf k},\omega) \rangle }{\delta h({\bf k},\omega)}
\right|_{h=0}.
\ee
This measures the response of the order-parameter to the source $h$.
Near equilibrium, the correlation and response functions 
are related though the fluctuation-dissipation relation \cite{FORSTER75},
%
%
\be
\label{EQ:FD}
C(\vk,\w) = \frac{2}{\w} \mbox{ Im } G(\vk,\w).
\ee

\subsection{The Kubo formula for the conductivity}

The linear a.c.\ conductivity, $\sigma (\w)$, for an isotropic material
can be defined in terms of the current response, ${\bf J}$
(which 
includes normal and supercurrent contributions), to an infinitesimal applied 
electric field, ${\bf E}$, through
%
%
\be
\label{EQ:CURRES}
{\bf J} (\w) = \sigma (\w) 
\mbox{ } {\bf E} (\w).
\ee
Since the quantities in Eq.\ (\ref{EQ:CURRES}) are evaluated at zero wavevector
we suppress their wavevector dependence. The conductivity is 
complex and has a real dissipative response, $\sigma'$,
and an imaginary reactive response, $\sigma''$:
%
%
\be
\label{EQ:REIM}
\sigma (\omega) = \sigma' (\omega) + i
\sigma'' (\omega). 
\ee

In linear response, the  conductivity is related 
to a current correlation function {\em via} the
Kubo formula \cite{KUBO85}.  
Near $T_{c}$ strong superconducting fluctuations give a  singular 
contribution to the conductivity which dominates the non-singular 
contribution due to normal electrons. Thus we may use the 
Kubo formula to calculate the real part of the conductivity 
due to superconducting fluctuations from the
supercurrent correlation function, evaluated at ${\bf E} = 0$
\cite{SCHMIDT68}: 
%
%
\be
\label{EQ:KUBO}
\sigma' (\omega) =  \frac{1}{2 d} 
\langle {\bf J}_{s} (\omega ) \cdot
{\bf J}_{s} (-\omega)  \rangle |_{ {\bf E} = 0}. 
\ee 
The supercurrent, ${\bf J}_{s}$ is
%
%
\be
\label{EQ:SUCUR}
{\bf J}_{s} ({\bf r},t) = - i e_{0}  ( \psi^{*} \nabla  
\psi - \psi \nabla \psi^{*}),
\ee
where $e_{0}$ is the bare charge of a Cooper pair.
The imaginary part of the conductivity can be obtained by applying the 
the Kramers-Kronig relations \cite{KUBO85} to Eq.\ (\ref{EQ:KUBO}).

The average in Eq.\ (\ref{EQ:KUBO}) is a four-point order-parameter 
average since ${\bf J}_{s}$ (\ref{EQ:SUCUR}) is quadratic in $\psi$.
Quite generally, 
this four-point average can be written as the sum of
 a ``disconnected'' product, $\sigma^{(2)}$, of two two-point averages,
 and a ``connected'' four point-average $\sigma^{(4)}$: 
%
%
\be
\label{EQ:CONDSPLIT}
\sigma' (\omega)  = 
\sigma^{(2)} (\omega) + \sigma^{(4)} (\omega)
\ee
with  
%
%
\be 
\label{EQ:DIS}
\sigma^{(2)} (\omega) = \frac{2 e^{2}_{0}}{d} 
\int_{k_{1} \w_{1}}^{\Lambda}
\mbox{ } k_{1}^{2} \mbox{ } C ({\bf k}_{1},\omega_{1}) C({\bf k}_{1}
, \omega_{1} + \omega) 
\ee
and
%
%
\be 
\label{EQ:VERTEX}
\sigma^{(4)} (\omega) = 
\frac{ 2 e^{2}_{0}}{d} \int_{k_{1} \w_{1} k_{2} \w_{2}}^{\Lambda}
 \mbox{ } {\bf k}_{1} \cdot {\bf k}_{2} \mbox{ }
C^{(4)}_{c}  ( {\bf k}_{1},\omega_{1},{\bf k}_{2},\omega_{2};\omega),
\ee
where the exact two-point order-parameter 
correlation function, $C ({\bf k},\omega)$, is defined in 
Eq.\ (\ref{EQ:OPCOR}) and
%
%
\bea
C^{(4)}_{c}  ( \vk_{1},\omega_{1},\vk_{2},\omega_{2};\omega)
 & \equiv & 
\nonumber \\ & &  \hspace{-1.2 in} 
\langle \psi (\vk_{1},\omega_{1}) 
\psi^{*} (\vk_{1},\omega_{1} - \omega) 
\psi (\vk_{2},\omega_{2}) \psi^{*} (\vk_{2},\omega_{2} + \omega) 
\rangle_{c}
\nonumber \\ & &
\label{EQ:FOURPT}
\eea
is the connected four-point order-parameter correlation function. 
%
%
\subsection{Iterative dynamic perturbation theory}
\label{SEC:PERTTH}

The order-parameter averages (\ref{EQ:OPCOR}) and  (\ref{EQ:FOURPT})
that appear in Eqs.\ (\ref{EQ:DIS}) and (\ref{EQ:VERTEX}) can  
be expanded as a perturbation series in
the bare non-linear coupling $u_{0}$ appearing in Eq.\ (\ref{EQ:FREE}).
Dynamic perturbation theory  for the time-dependent Ginzburg Landau
equation (\ref{EQ:GLMOD}) can be implemented either by using a 
Martin-Siggia-Rose \cite{MSR,DEDOMINICIS78} field-theoretical formalism, 
or by a direct 
iteration of the equation of motion \cite{MA76,MAZENKO}.
 The iterative approach involves less formal machinery and
will be used here.
%
%
\begin{figure}
\begin{center}
    \leavevmode
    \epsfxsize=3.25 in
    \epsffile{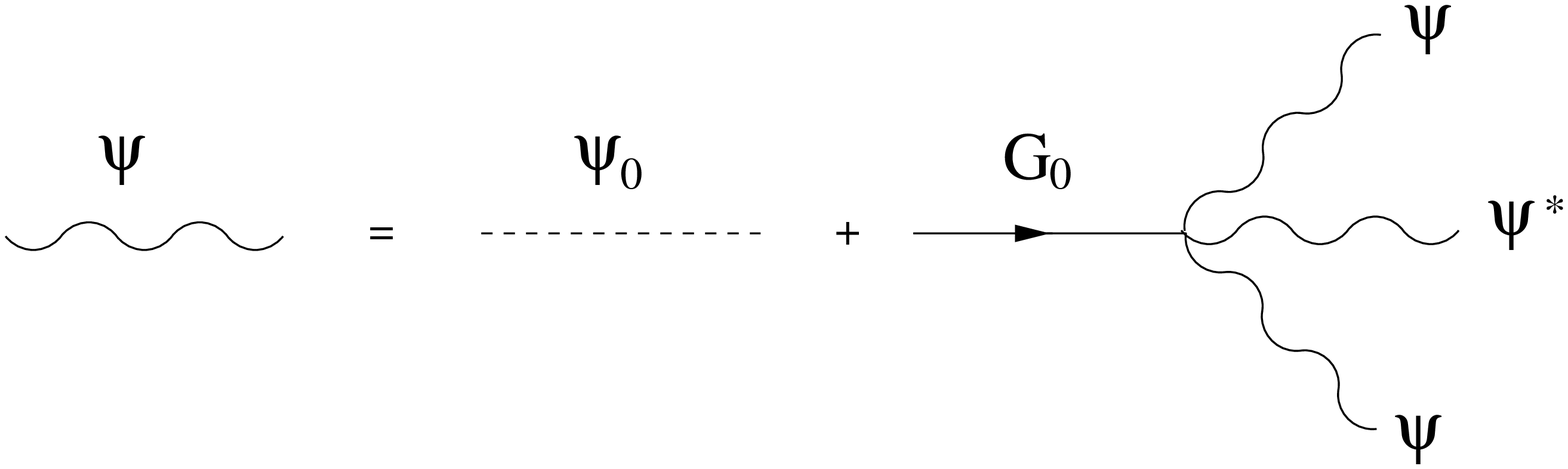}
  \end{center}

{\small
FIG. 2. The diagrammatic representation of the equation of motion 
(\ref{EQ:GLIT}). Wiggly lines correspond to the order-parameter $\psi$
(a starred wiggly line is $\psi^{*}$).
The dotted line represents the Gaussian field $\psi_{0}$. 
The Gaussian response function $G_{0}$ (\ref{EQ:RESG})
is shown as a line with an arrow. The vertex, where
the response function meets three wiggly lines contains a factor $-u_{0}$,
as well as $V$
(\ref{EQ:V}), which conserves wavevector and frequency at the vertex.
Iteration corresponds to replacing the wiggly lines on the right-hand  
side with either the first or second term on the right-hand side. In this way,
one generates a series in $u_{0}$.}
\label{FIG:ITER}
\end{figure}

The equation of motion (\ref{EQ:GLMOD}) can be explicitly written in  
Fourier space as
%
%
\bea
\psi(\vk, \omega) & = & \psi_{0} (\vk, \omega) 
-  u_{0} \mbox{ }  G_{0} (\vk,\omega)  
\nonumber \\ & & 
\times \int_{k_{1} \w_{1} k_{2} \w_{2} k_{3} \w_{3}}^{\Lambda} \hspace{-.3 in}
V \times \mbox{ } 
\psi(\vk_{1},\omega_{1}) \psi^{*}(\vk_{2},\omega_{2}) 
\psi(\vk_{3},\omega_{3}),
\nonumber \\ & &
\label{EQ:GLIT}
\eea
where
%
%
\be
\label{EQ:OPG}
\psi_{0} (\vk, \omega ) = 
\frac{1}{\Gamma_{0}} G_{0} (\vk, \omega) \mbox{ } \zeta (\vk, 
\omega),
\ee
%
%
\be
\label{EQ:RESG}
G_{0} (\vk, \omega)  =  \left(- \frac{i \omega }{\Gamma_{0}} + 
r_{0} + k^{2} \right)^{-1}
\ee
and
%
%
\be
\label{EQ:V}
V = 
(2 \pi)^{d+1}  \delta(\vk - \vk_{1} + \vk_{2} - \vk_{3} ) 
\delta(\omega - \omega_{1} + \omega_{2} - \omega_{3} ).
\ee

The Gaussian theory neglects the non-linear interaction ($u_{0} = 0$).
In this case Eq.\ (\ref{EQ:GLIT}) reduces to  $\psi = \psi_{0}$, and 
the order-parameter is a Gaussian field by virtue of Eq.\ (\ref{EQ:OPG}) and
the fact that $\zeta$ is Gaussian. The order-parameter correlation 
function (\ref{EQ:OPCOR}) can be evaluated using Eq.\ (\ref{EQ:NOISE})
and is
%
%
\bea
C_{0} ({\bf k},\omega) & \equiv & \langle \psi_{0} ({\bf k},\omega)
\psi_{0}^{*} ({\bf k},\omega) \rangle  
\nonumber \\
& = & \frac{2 \Gamma_{0}}{\omega^{2} + 
\Gamma_{0}^{2} (r_{0} + k^{2} )^{2} }.
\label{EQ:CORG}
\eea
If a term coming from $F_{h}$ (\ref{EQ:SOURCE}) is included in the 
equation of motion (\ref{EQ:GLIT}), it is straightforward to show that $G_{0}$
(\ref{EQ:RESG}) is the Gaussian response function.
 A glance at Eqs.\ (\ref{EQ:RESG})
and (\ref{EQ:CORG}) shows that the Gaussian theory satisfies the 
fluctuation-dissipation relation, Eq.\ (\ref{EQ:FD}).

Since $\psi$ appears in the integral on the right-hand side of 
Eq.\ (\ref{EQ:GLIT}), this equation 
can be iterated to produce an expansion for $\psi$ in powers of the 
bare coupling constant $u_{0}$. Averages containing 
$\psi$ are then expressed as sums of higher-point
Gaussian averages  over $\psi_{0}$, which break up into products of $C_{0}$'s.
To keep track of the algebra, it is helpful to use the graphical 
representation of Eq.\ (\ref{EQ:GLIT}) shown in Fig.\ 2.
In the graphical context
iteration corresponds to ``putting branches on the tree'' and
averaging corresponds to joining two conjugate dashed lines ($\psi_{0}$)
to form a correlation
function $C_{0}$. By examining all possibilities for joining for a given 
average, a series of graphs is generated with the proper symmetry 
factors. In dynamical perturbation 
theory there are two propagators: the response function $G_{0}$, denoted
by an arrow, and the correlation function $C_{0}$, denoted by a line with 
a circle on it. Wavevector and frequency are assigned to these lines on 
the basis of conservation of wavevector and frequency
at the graph vertices, given by V in Eq.\
(\ref{EQ:V}). Wavevectors and frequencies flowing around loops are integrated 
over. More details of the graph rules can be found in \cite{MA76,MAZENKO}.
An example of this procedure is the self-energy diagram, Fig.\ 
3, and the corresponding algebraic expression 
(\ref{EQ:SELF}).
%
%
\subsection{Renormalization of the theory and the XY fixed-point}
\label{SEC:MULTREN}


It is computationally convenient to dimensionally regularize the theory 
and renormalize {\em via} minimal subtraction \cite{DEDOMINICIS78,AMIT}.
This will produce a sensible $\e = 4 - d$ expansion.
To be more concrete, we define the renormalized ``coupling constant,''
$u$, 
in terms of the bare coupling constant, $u_{0}$, by
%
%
\be
u \equiv Z_{u} \mbox{ } u_{0},
\ee
and define the dimensionless, renormalized coupling constant, 
$\bar{u}$ as 
%
%
\be
\label{EQ:DIMCOUP}
\bar{u} \equiv \frac{S_{d}}{2 (2 \pi)^{d}} \mbox{ } u \kappa^{-\e},
\ee
where $\kappa$  is an arbitrary wavevector scale and 
 $S_{d}$ is the surface area of the unit sphere in $d$ dimensions.
The renormalization constant  
$Z_{u} = 1 + O(\bar{u})$ \cite{AMIT}. Since only $\bar{u}^{2}$ will appear
in the conductivity, and we neglect terms of $O(\bar{u}^{3})$ and higher,
we may approximate $Z_{u} = 1$. 

Renormalization of the bare response function (\ref{EQ:RES}) 
provides the remaining renormalization
constants. The bare inverse response function including self-energy 
corrections, $\Sigma$, may be written
%
%
\be
\label{EQ:DYSON}
G^{-1} (\vk, \w) = G^{-1}_{0} (\vk, \w) - \Sigma(\vk, \w).
\ee
The renormalized inverse response function $G^{-1}_{R} (\vk,\w)$ may be 
expressed in terms of the bare quantity (\ref{EQ:DYSON}) by
%
%
\be
\label{EQ:RENRES}
G^{-1}_{R} (\vk,\w) \equiv Z_{\psi} \mbox{ } G^{-1} (\vk, \w),
\ee
where the renormalization constant $Z_{\psi}$ comes from ``wavefunction''
renormalization (a rescaling of $\psi$) 
and, in the minimal subtraction scheme, is given by 
\cite{AMIT,BETADEFN}
%
%
\be
\label{EQ:WFREN}
Z_{\psi} = 1 - \frac{1}{\e} \bar{u}^{2} + O(\bar{u}^{3}).
\ee
The renormalized ``mass'' $r$ is defined as
\be
\label{EQ:ALRES}
r \equiv G_{R}^{-1} (0,0),
\ee
which, using Eqs.\ (\ref{EQ:RESG}), (\ref{EQ:DYSON}) and (\ref{EQ:RENRES}),
 is related to the bare mass $r_{0}$ by
%
%
\be
\label{EQ:REALPHA}
r = Z_{\psi} [ r_{0} - \Sigma(0, 0)].
\ee
Near $T_{c}$ the physical response function at zero wavenumber and 
frequency behaves as $G_{R} (0,0) = \xi^{2-\eta} \kappa^{-\eta}$, where
$\eta$ is the usual correlation function exponent and $\xi$ is
the order-parameter correlation length which diverges as 
%
%
\be
\label{EQ:CRITXI}
\xi \sim |T-T_{c}|^{-\nu},
\ee
with the critical exponent $\nu$. Thus, from (\ref{EQ:ALRES}),
we have
%
%
\be
\label{EQ:ALXI}
r = \xi^{-2 + \eta} \kappa^{\eta}.
\ee
Since we are neglecting magnetic fluctuations and working at the 
``uncharged'' fixed point, the renormalized charge, $e$, is simply
the bare charge: $e=e_{0}$.
Finally, the bare relaxation rate $\Gamma_{0}$, 
appearing in the dynamic response 
function (\ref{EQ:DYSON}) is related to the renormalized relaxation 
rate $\Gamma$ by
%
%
\be
\label{EQ:GAMREN}
\frac{1}{\Gamma_{0}} = Z_{\Gamma} \mbox{ } \frac{1}{\Gamma},
\ee
where, from minimal 
subtraction, the renormalization constant $Z_{\Gamma}$ for this relaxational
model is 
\cite{DEDOMINICIS78,BETADEFN}
%
%
\be
\label{EQ:ZGAM}
Z_{\Gamma} = 1 - \frac{c}{\e} \bar{u}^{2} 
+ O(\bar{u}^{3}).
\ee
The constant $c$ is given by Eq.\ (\ref{EQ:C}). 


Near $T_{c}$, as one probes the long-wavelength physics,
the coupling $\bar{u}$ flows towards the fixed point value $\bar{u}^{*}$
determined by the IR-stable zeros of the renormalization-group beta function
$\beta(\bar{u}^{*}) = 0$ \cite{AMIT}. This mechanism is responsible
for  universality. To leading order in the $\e$-expansion, $\bar{u}^{*}$ is
\cite{AMIT,BETADEFN}
%
%
\be
\label{EQ:BFIXED}
\bar{u}^{*} = \frac{\e}{10} + O(\e^{2}).
\ee
This is the Wilson-Fisher \cite{WILSON72} fixed-point for the XY-model.
The correlation function exponent $\eta$ is related to $Z_{\psi}$,
(\ref{EQ:WFREN}) and has the following expansion in $\bar{u}^{*}$:
%
%
\be
\label{EQ:B-ETA}
\eta= 2 \mbox{ } (\bar{u}^{*})^{2} + O ( (\bar{u}^{*})^{3}).
\ee
The result $\nu \approx 2/3$ quoted in the Introduction, 
which also appears in (\ref{EQ:CRITXI}), is 
an extrapolation of the $\e$-expansion result to three dimensions. 
Finally, the dynamic exponent $z$ is related to $Z_{\Gamma}$ 
(\ref{EQ:ZGAM}) for the relaxational dynamics, 
and given by $z=2 + c \eta$ with $c = 6 \ln 4/3 -1$ and
$\eta$ given by  Eq.\ (\ref{EQ:B-ETA}).
Thus, by reorganizing the theory as an expansion in $\e$
and using the fixed point value $\bar{u}^{*}$ for the coupling,
the IR divergences near criticality can be sensibly treated and lead 
to corrections to the Gaussian exponents. 

Even after we renormalize the conductivity as described above,
some poles in $\e$ will remain.
These poles are due to UV-divergences in the theory for the conductivity 
that appear even at the
Gaussian level and have nothing to do with the critical behaviour. 
These poles must be eliminated by adding a constant to
the conductivity, as will be discussed in Sec.\ \ref{SEC:RENORM}. 

%
%
\section{The conductivity in the Gaussian approximation}

We now review earlier work on the a.c.\ conductivity involving non-interacting,
Gaussian fluctuations \cite{SCHMIDT68,DORSEY91}, and set $u_{0}=0$
in Eq.\ (\ref{EQ:FREE}). 
In the Gaussian approximation the connected piece  of
the conductivity, Eq.\ (\ref{EQ:VERTEX}), is zero.  
Thus, from Eqs.\ (\ref{EQ:CONDSPLIT}) and (\ref{EQ:DIS})
one has 
%
%
\be
\label{EQ:CONDG}
\sigma' (\omega) = \frac{2 e_{0}^{2}}{d} \int_{k_{1} \w_{1}}^{\Lambda}
\mbox{ } 
k_{1}^{2} \mbox{ } C_{0} ({\bf k}_{1},\omega_{1}) C_{0} 
({\bf k}_{1}, \omega_{1} + \omega),
\ee
where $C_{0}$ is given by (\ref{EQ:CORG}).
The calculation of the integral in Eq.\ (\ref{EQ:CONDG}) involves a contour
integration over the frequency variable, and then a straightforward
 evaluation of the remaining wavevector integral, with the cutoff $\Lambda$
set to infinity. The complex conductivity takes the form
\cite{SCHMIDT68,DORSEY91}:
%
%
\be
\label{EQ:GSIG}
\sigma (\omega) = \frac{e_{0}^{2}}{2 \Gamma_{0}} \bar{\sigma}  
\mbox{ }
\frac{\xi_{0}^{4-d}}{4-d} 
\mbox{ } S_{G} (y_{0}),
\ee
where
%
%
\be 
\label{EQ:BSIG}
\bar{\sigma}  = \frac{S_{d}}{(2 \pi)^{d}} \Gamma(d/2) \Gamma(3-d/2)
\ee
and the scaled frequency $y_{0}$ is 
%
%
\be
\label{EQ:Y0}
y_{0} = \frac{\omega \xi_{0}^{2}}{2 \Gamma_{0}}.
\ee
The Gaussian 
order-parameter correlation length $\xi_{0}$ is defined as
%
%
\be
\label{EQ:CORLEN}
\xi_{0} \equiv r_{0}^{-1/2},
\ee
thus $\xi_{0} \sim |T - T_{c0} |^{-1/2}$ and 
$\nu = 1/2$ in the Gaussian theory.
The real part of the scaling form 
$S_{G}$ is computed from Eq.\ (\ref{EQ:CONDG}) to be
%
%
\be
\label{EQ:CGR}
S'_{G} (y_{0}) = \frac{8}{d (d - 2)}
\frac{1}{y_{0}^{2}} \left[ 
1 - (1 +  y_{0}^{2} )^{d/4} \cos \left(\frac{d}{2} \tan^{-1} y_{0}
\right) \right].
\ee
The imaginary part of the
conductivity is obtained from Eq.\ (\ref{EQ:CGR}) using the Kramers-Kronig
relations. The result for the complex scaling form is then 
\be
\label{EQ:CGI}
S_{G} (y_{0}) = \frac{8}{d (d - 2)} 
\frac{1}{y_{0}^{2}} \left[ 1 - \frac{d}{2} \mbox{ } i y_{0} 
- (1 - i y_{0})^{d/2} \right].
\ee
The Gaussian result, 
Eq.\ (\ref{EQ:GSIG}) with the definition (\ref{EQ:Y0}), 
satisfies the FFH hypothesis
(\ref{EQ:SCALCON}) with $z=2$.

We note two properties of these results that will be
important  later. The first is that 
Eq.\ (\ref{EQ:GSIG}) has a factor of $\e = 4 - d$ in the denominator. This is 
a consequence of setting the cutoff $\Lambda$ to infinity, and indicates
that even the Gaussian theory is sensitive to the cutoff in four dimensions.
The second property is that $S'_{G}$ (\ref{EQ:CGR}) has the $\e$-expansion
%
%
\be
\label{EQ:EEXPG}
S'_{G} (y_{0}) = 1 + \sum_{i=1}^{\infty} \e^{i} \mbox{ } S_{i} (y_{0}).
\ee
The coefficient of $\e$ in Eq.\ (\ref{EQ:EEXPG}),
\be
\label{EQ:S1}
S_{1} (y_{0}) =  \frac{3}{4} + \frac{1}{4 y_{0}^{2}} [ (1 - y_{0}^{2}) 
\ln (1 + y_{0}^{2}) 
- 4 y_{0} \tan^{-1} y_{0} ] ,
\ee
is interesting because it appears later in both the disconnected and
the connected pieces of the conductivity.
%
%
\section{Disconnected piece of the conductivity}

To go  beyond the  Gaussian theory  requires the calculation of both the full
two-point correlation 
function (\ref{EQ:OPCOR}), including self-energy corrections, 
and the four-point average (\ref{EQ:FOURPT}) which appear in the
conductivity through Eqs.\ (\ref{EQ:DIS}) and (\ref{EQ:VERTEX}).  
The calculations must be performed to $O(u^{2})$, where
the first corrections to the Gaussian result $z=2$ occur.
In this section we examine the disconnected piece of the 
conductivity (\ref{EQ:DIS}). The next section tackles the connected 
piece.

We first dimensionally regularize and renormalize the theory 
as outlined in 
Sec.\ \ref{SEC:MULTREN}.
From Eq.\ (\ref{EQ:DIS}),
the disconnected contribution to the conductivity is then
%
%
\be
\label{EQ:DIS2}
\sigma^{(2)}(\omega) = \frac{2 e^{2}}{d}
\int_{k_{1} \w_{1}} k_{1}^{2} 
\mbox{ } 
C(\vk_{1},\w_{1}) C(\vk_{1},\w_{1} + \w),
\ee
where $C$ is the full correlation function (\ref{EQ:OPCOR}), including 
self-energy corrections.
We will calculate the response function $G$ (\ref{EQ:RES}) to $O(u^{2})$ 
and use the fluctuation-dissipation relation (\ref{EQ:FD})
to get $C$. With the definition (\ref{EQ:REALPHA})
of the renormalized mass, $r$, the inverse response function (\ref{EQ:DYSON})
may be written as
%
%
\be
\label{EQ:IRES}
G^{-1}(\vk,\w) = G^{-1}_{0} (\vk,\w) - [ \Sigma (\vk,\w) - \Sigma (0,0) ],
\ee
where now 
\bea
\label{EQ:SUB1}
r_{0} & \rightarrow & r/Z_{\psi} \\
\label{EQ:SUB2}
\Gamma_{0} & \rightarrow & \Gamma/Z_{\Gamma}
\eea
in $G_{0}$ (\ref{EQ:RESG}) and $C_{0}$ (\ref{EQ:CORG}).
To $O(u^{2})$ only the ``Saturn'' diagram $\Sigma_{s} (\vk,\w)$ 
shown in Fig.\ 3
contributes to Eq.\ (\ref{EQ:IRES}) since,
to this order, it is the only piece of the self-energy 
that is wavevector- and
frequency-dependent. Applying the rules outlined in Sec.\ \ref{SEC:PERTTH}
to  Fig.\ 3 gives 
%
%
\bea
\Sigma_{s} (\vk,\omega) & = & 6 u^{2} \int_{k_{2} \w_{2} k_{3} \w_{3} }
C_{0} (\vk_{2},\w_{2}) C_{0} (\vk_{3},\w_{3}) 
\nonumber \\ & & \hspace{.7 in}
\times G_{0} (\vk - \vk_{2} 
-\vk_{3},\w - \w_{2} - w_{3}).
\nonumber \\ & & 
\label{EQ:SELF}
\eea
%
%
\begin{figure}
\begin{center}
    \leavevmode
    \epsfxsize=1.5 in
    \epsffile{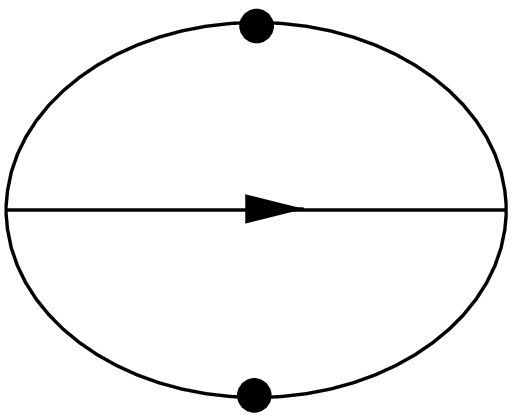}
  \end{center}

{\small FIG. 3. The Saturn diagram, $\Sigma_{s}$, for the self-energy
consists of two loops formed by two correlation functions 
$C_{0}$ (lines with circles)
and one response function $G_{0}$ (line with an arrow). Wavevector and
frequency flow through the diagram in accordance with the discussion 
in Sec.\ \ref{SEC:PERTTH}. \hspace{3 in}}
\label{FIG:SELF}
\end{figure}
\hspace{-.2 in}
The correlation function $C$ is then obtained from (\ref{EQ:FD}) and 
(\ref{EQ:IRES}):
%
%
\bea
C(\vk,\w) & = & C_{0} (\vk,\w) 
\nonumber \\ & & \hspace{-.1 in}
+ \frac{2}{\w} \mbox{ Im} 
\{ G_{0}^{2} (\vk,\w) [\Sigma_{s} (\vk,\w) - \Sigma_{s} (0,0) ] \} + O(u^{3}).
\nonumber \\ & &
\label{EQ:CTOB2}
\eea
Thus the  disconnected piece of the conductivity (\ref{EQ:DIS2})
can be expressed in terms
of the integrals
\be
I_{1} (\w) =  \frac{2 e^{2}}{d} \int_{k_{1} \w_{1}} k_{1}^{2} \mbox{ }
  C_{0} (\vk_{1},\w_{1})  
C_{0} (\vk_{1},\w_{1} + \w)  
\label{EQ:I1}
\ee
and
\bea
I_{2} (\w)  & =  &  \frac{4 e^{2}}{d} \mbox{Im} 
\int_{k_{1} \w_{1}} k_{1}^{2} \mbox{ }  C_{0} (\vk_{1},\w_{1})  
\frac{1}{\w_{1} + \w} 
G_{0}^{2} (\vk_{1},\w_{1} +\w) 
\nonumber \\ & & \hspace{ .7 in}
 \times [ \Sigma_{s} (\vk_{1},\w_{1}+\w) - \Sigma_{s} (0,0) ],
\label{EQ:I2}
\eea
by writing
\be
\sigma^{(2)}(\omega) = I_{1} (\w) + 2 I_{2} (\w) + O(u^{3}).
\ee
Each integral is dealt with separately below.
%
%
\subsection{The integral $I_{1}$}
 
The only differences between $I_{1}$ (\ref{EQ:I1}) 
and the starting point (\ref{EQ:CONDG}) of the Gaussian calculation 
are the substitutions: Eqs.\  (\ref{EQ:SUB1}), (\ref{EQ:SUB2}) and
$e_{0} \rightarrow e$.     
 Transcribing the real part of the Gaussian result (\ref{EQ:GSIG}) gives
%
%
\be
\label{EQ:I1EVAL}
I_{1} ( \w) = \frac{e^{2}}{2 \Gamma} \bar{\sigma} \kappa^{-\e}
\mbox{ }  
\frac{x^{-\e}}{\e} Z_{\psi}^{\e /2} Z_{\Gamma} 
\mbox{ } S'_{G}(\tilde{y}),
\ee
with $S'_{G}$ given in Eq.\ (\ref{EQ:CGR}) and 
\be
\tilde{y} = \frac{\w Z_{\psi} Z_{\Gamma}}{2 \Gamma r}.
\ee
The dimensionless measure of the nearness to the 
transition is
\be
x = \frac{\sqrt{r}}{\kappa},
\ee
where the arbitrary wavevector scale $\kappa$ was introduced earlier in 
Eq.\ (\ref{EQ:DIMCOUP}).
From the expression (\ref{EQ:ALXI}) for $r$ we have
\be
\label{EQ:XTEMP}
x = (\xi \kappa)^{-1 + \eta/2}.
\ee
The function $S'_{G}(\tilde{y})$ can be expressed in terms of the 
scaled frequency $y$, 
%
%
\be
\label{EQ:Y}
y = \frac{\omega \xi^{z} \kappa^{z-2}}{2 \Gamma}
\ee
with $z$ given by Eq.\ (\ref{EQ:ZED}), by the expansion
%
%
\bea
S'_{G}(\tilde{y}) & = & S'_{G}(y) + \partial_{y} S'_{G}(y) \mbox{ } 
(\tilde{y} - y) + 
\frac{1}{2} \partial_{y}^{2} S'_{G}(y) \mbox{ } (\tilde{y} - y)^{2} 
\nonumber \\ & & \hspace{.75 in} 
+ 
\frac{1}{2} \partial_{y}^{2} S'_{G}(y) \mbox{ } (\tilde{y} - y)^{2}
+ \cdots,
\label{EQ:SGT}
\eea
where $\partial_{y}$ indicates a derivative with respect to $y$.
The results (\ref{EQ:WFREN}) for $Z_{\psi}$,
(\ref{EQ:ALXI}) for $r$
and (\ref{EQ:ZGAM}) for $Z_{\Gamma}$ are used to obtain the 
following relation between $\tilde{y}$ and $y$:
%
%
\be
\label{EQ:TYY}
\tilde{y} - y = y (c+1) \left( \eta \ln x - \frac{1}{\e} \bar{u}^{2}
\right) 
+ O(\bar{u}^{2} \e),
\ee
where $c$ is given by Eq.\ (\ref{EQ:C}). Using
equation (\ref{EQ:TYY}), the expansion (\ref{EQ:EEXPG}) of $S'_{G}$, 
and the fact that $\eta$ (\ref{EQ:B-ETA}) is $O(\e^{2})$,
we write (\ref{EQ:SGT}) as
\bea
S'_{G}(\tilde{y}) & = & 1 + \e S_{1} (y) + \e^{2} S_{2} (y) +  \e^{3} S_{3}(y)
\nonumber \\ & & 
+ \e (c+1) \left(\eta \ln x - \frac{1}{\e} \bar{u}^{2} \right) 
[y \partial_{y} S_{1} (y) + \e y \partial_{y} S_{2} (y) ] 
\nonumber \\ & & \hspace{ 1.5 in} 
+ O(\bar{u}^{2} \e^{2},\e^{4}).
\label{EQ:SGEE}
\eea  
We now use the expansions
(\ref{EQ:WFREN}) of $Z_{\psi}$ and (\ref{EQ:ZGAM})
of  $Z_{\Gamma}$ together with Eq.\ (\ref{EQ:SGEE}) to 
write $I_{1}$ (\ref{EQ:I1EVAL}) as a series in $\bar{u}$, with 
coefficients expanded in powers of $\e$. Terms of $O(\bar{u}^{2} \e,\e^{3})$ 
and higher are neglected (since the fixed-point value $\bar{u}^{*}$ 
(\ref{EQ:BFIXED}) is $O(\e)$ we are effectively working to $O(\e^{2})$).
The result, written in a form that will  be convenient for 
later analysis, is
\end{multicols}
%
%
\bea
I_{1}(\w) & = & \frac{e^{2}}{2 \Gamma} \bar{\sigma} \kappa^{-\e} 
 \left(1 - \frac{1}{2} \bar{u}^{2} \right) \left\{
- \frac{c}{\e^{2}} \bar{u}^{2}  + \frac{1}{\e} + \frac{c}{\e} \bar{u}^{2} \ln x
- \frac{c}{\e} \bar{u}^{2} S_{1} (y)  
- \frac{c+1}{\e}  \bar{u}^{2} \mbox{ } y \partial_{y} S_{1} (y)
- \ln x +  \frac{\e - c \bar{u}^{2}}{2} (\ln x)^{2} 
\nonumber \right. \\ & & \hspace{.5 in}
- \frac{\e^{2}}{6}
 (\ln x)^{3}   
+ \left[ 1 - (\e - c \bar{u}^{2}) \ln x + \frac{\e^{2}}{2} (\ln x)^{2}
\right] \mbox{ } S_{1}(y)
+ (\e- c \bar{u}^{2}) (1 - \e \ln x) \mbox{ } S_{2} (y)
+ \e^{2} S_{3}(y)
\nonumber \\ & & \hspace{2 in} 
+ (c+1) (\eta + \bar{u}^{2}) \mbox{ } y \partial_{y} S_{1}(y) 
\mbox{ } \ln x - (c+1) \bar{u}^{2} \mbox{ }  y \partial_{y} S_{2} (y)
+ O(\bar{u}^{2} \e,\e^{3}) \bigg{\}}.
\label{EQ:I1FIN}
\eea
%
%
\subsection{The integral $I_{2}$}
\label{SEC:DIS2}

The calculation of $I_{2}$, Eq.\ (\ref{EQ:I2}), is involved so we
only outline it here. The first step is to re-scale the internal
wavevectors  and frequencies in Eq.\ (\ref{EQ:I2}) by 
%
%
\bea
\label{EQ:SCALK}
\vk_{i} & \rightarrow & \sqrt{r} \mbox{ }  \vk_{i} \\
\label{EQ:SCALW}
\w_{i} & \rightarrow & \Gamma r \mbox{ } \w_{i}, 
\eea
where $i$ = 1, 2, 3 (remember that $\Sigma_{s}$ contains an integral over
$\vk_{2} \w_{2} \vk_{3} \w_{3}$), and write
\be
\label{EQ:I2SCALED}
I_{2} (\w) = \frac{12 e^{2}}{d \Gamma} \kappa^{-\e} \mbox{ }  
(u \kappa^{-\e})^{2} \mbox{ } x^{-3 \e} \tilde{I}_{2} (y).
\ee
The dimensionless integral in Eq.\ (\ref{EQ:I2SCALED}), 
\bea
\tilde{I}_{2} (y) & = & 16 \mbox{ Im} 
\int_{k_{1} \w_{1} k_{2} \w_{2} k_{3} \w_{3}} 
k_{1}^{2} \mbox{ } \tilde{C}_{0} (\vk_{1},\w_{1})
\frac{1}{\w_{1} + 2 y} \tilde{G}_{0}^{2} (\vk_{1},\w_{1}+2 y) 
\tilde{C}_{0} (\vk_{2}, \w_{2}) \tilde{C}_{0} (\vk_{3}, \w_{3}) 
\nonumber \\ & & \hspace{2 in} 
\times
[ \tilde{G}_{0} (\vk_{1} -\vk_{2} -\vk_{3}, 2 y + \w_{1} - \w_{2} - \w_{3}) 
- \tilde{G}_{0} (-\vk_{2} -\vk_{3}, - \w_{2} - \w_{3})],
\label{EQ:TI2}
\eea
is written in terms of the dimensionless functions
\be
\label{EQ:TG0}
\tilde{G}_{0} (\vk,\w) =  \frac{1}{-i \w + 1 + k^{2}} 
\ee
and
\be
\label{EQ:TC0}
\tilde{C}_{0} (\vk,\w) =  \frac{1}{\w^2 + (1 + k^{2})^{2}}.
\ee
Since $I_{2}$ is already $O(u^{2})$ we have simply replaced all bare
coefficients in Eq.\ (\ref{EQ:I2SCALED}) by renormalized ones, and used
the scaled frequency $y$ from Eq.\ (\ref{EQ:Y}).

The second step is to evaluate the three frequency integrals in
Eq.\ (\ref{EQ:TI2}) by contour integration. The calculation is straightforward 
and yields
%
%
\be
\label{EQ:I22}
\tilde{I}_{2} (y) = \mbox{ Re } [\tilde{I}_{2}^{a} (y) + 
\tilde{I}_{2}^{b} (y) + 
\tilde{I}_{2}^{c} (y) ],
\ee
with 
\bea
\tilde{I}_{2}^{a} (y) & = & \int_{0}^{1} d v \mbox{ } (1 - v)
\int_{k_{1} k_{2} k_{3}} k_{1}^{2} \mbox{ } \left[ \frac{2}{a_{1}^{3}}
- \frac{1}{(a_{1} + i y v)^{3}} 
\right] \frac{1}{a_{2} a_{3} (a_{2} + a_{3} 
+ a_{4}) (a_{5} + 2 i y)} ,
\label{EQ:I2a} \\ 
\tilde{I}_{2}^{b} (y) & = & 3 \int_{0}^{1} d v \mbox{ } (1 - v)
\int_{k_{1} k_{2} k_{3}} k_{1}^{2} \mbox{ }
\frac{1}{(a_{1} + i y v)^{4} a_{2} a_{3} } 
\left[ \frac{1}{a_{2} + a_{3} + a_{4}} - \frac{1}{a_{2} + a_{3} + \bar{a}_{4}}
\right] ,
\\
\tilde{I}_{2}^{c} (y)& = & - 3 i  y \int_{0}^{1} d v 
\mbox{ } (1 - v)
\int_{k_{1} k_{2} k_{3}} k_{1}^{2} \mbox{ }
\frac{1}{(a_{1} + i y v)^{4} a_{2} a_{3} (a_{2} + a_{3} + a_{4})
(a_{5} + 2 i y)}, 
\label{EQ:I2c}
\eea
where, for convenience, we define
\bea
a_{i} & \equiv & 1 + k_{i}^{2}, \hspace{1 in} \mbox{ $i$ = 1, 2, 3}\\
a_{4} & \equiv & 1 + (\vk_{1} + \vk_{2} + \vk_{3})^{2}, \\
\bar{a}_{4} & \equiv & 1 + (\vk_{2} + \vk_{3})^{2}, \\
a_{5} & \equiv & a_{1} + a_{2} + a_{3} + a_{4}.
\eea
Note that we have used the Feynman formula
%
%
\be
\label{EQ:FEYFORM}
\frac{1}{c_{1}^{\alpha_{1}} c_{2}^{\alpha_{2}}} =
\frac{\Gamma(\alpha_{1} + \alpha_{2})}{\Gamma(\alpha_{1}) \Gamma(\alpha_{2})}
\int_{0}^{1} d v (1-v)^{\alpha_{1}-1} v^{\alpha_{2} -1} 
\frac{1}{[(1-v) c_{1} + v c_{2} ]^{\alpha_{1} + \alpha_{2}} }
\ee
with the Feynman parameter $v$ to group and simplify terms in Eqs.\
(\ref{EQ:I2a}-\ref{EQ:I2c}).

The final step is to evaluate the wavevector integrals in Eqs.\
(\ref{EQ:I2a}-\ref{EQ:I2c}) using (\ref{EQ:FEYFORM}) and $\e$-expand  
the resulting integrals over Feynman parameters to $O(\e^{0})$. 
An example of this procedure appears in Appendix \ref{APPENDIX}. 
The results for Eqs.\ (\ref{EQ:I2a}-\ref{EQ:I2c}) are
%
%
\bea
\label{EQ:I2aE}
 \tilde{I}_{2}^{a} (y) & = & A_{d} \left[ \frac{1}{6 \e^{2}} 
\ln \frac{4}{3} + \frac{1}{2 \e} f_{1} (y) \ln \frac{4}{3} 
 + \frac{0.003}{\e}
 + {\cal F}_{2}^{a} (y) + O(\e) \right] \\
\tilde{I}_{2}^{b} (y) & = &  A_{d} \left[ - \frac{1}{12 \e^{2}} 
+ \frac{1}{ 4 \e} f_{1} (y) - \frac{0.104}{\e} + {\cal F}_{2}^{b} (y) 
+ O(\e) \right] \\
\tilde{I}_{2}^{c} (y) & = & A_{d} \left[ - \frac{1}{2 \e} f_{2} (y)
\ln \frac{4}{3}+ {\cal F}_{2}^{c} (y) + O(\e) \right],
\label{EQ:I2cE}
\eea
where ${\cal F}_{2}^{a}$, ${\cal F}_{2}^{b}$ and ${\cal F}_{2}^{c}$
 are $O(\e^{0})$ functions of $y$ which we do not need to determine,
%
%
\be
\label{EQ:AD}
A_{d} = \left(\frac{S_{d}}{2(2 \pi)^{d}} \right)^{3} 
[ \Gamma(2 - \e/2) ]^{2} \Gamma(3 - \e/2)
\Gamma(1 + 3 \e /2),
\ee
and 
%
%
\bea
\label{EQ:F1}
f_{1}(y) & = & \int_{0}^{1} d v \mbox{ } (1-v) \mbox{ } \ln (1 + i y v) \\
\label{EQ:F2}
f_{2} (y) & = & i y \mbox { } \int_{0}^{1} d v \frac{1-v}{1 + i y v}.
\eea
It is straightforward to show that
%
%
\bea
\label{EQ:FS1}
\mbox{Re } [f_{1}(y)] & = & - S_{1} (y) \\
\mbox{Re } [f_{2}(y)] & = & -2 S_{1} (y) - y \partial_{y} S_{1} (y),
\eea
where $S_{1}$ was defined in (\ref{EQ:S1}). We use this result, along with
Eqs.\ (\ref{EQ:I22}) and (\ref{EQ:I2aE}-\ref{EQ:I2cE}) 
to write $I_{2}$ (\ref{EQ:I2SCALED}) as a product of $\bar{u}^{2}$ and
a series in $\e$. In particular, we have
\bea
2 I_{2} (\w) &
= & \frac{e^{2}}{2 \Gamma} \bar{\sigma} \kappa^{-\e} \mbox{ }
\bar{u}^{2} \Big{[} \frac{c-2}{3 \e^{2}} - \frac{c-2}{\e} \ln x 
+ \frac{c-2}{\e} S_{1}(y) + \frac{(c+1)}{\e} y \partial_{y} S_{1} (y) 
- \frac{0.787}{\e} 
+ 2.36 \ln x
+ \frac{3(c-2)}{2} (\ln x)^{2} 
 \nonumber \\ & & 
 \nonumber \\ & & \hspace{2 in}
- 3 (c-2) S_{1}(y)  \mbox{ } \ln x
 - 3 (c+1) \mbox{ } y \partial_{y} S_{1} (y) 
\mbox{ } \ln x + {\cal D} (y)  + O(\e) \Big{]},
\label{EQ:I2FIN}
\eea
where $c$ is given in Eq.\ (\ref{EQ:C}) and
\be
{\cal D} (y) = 0.233 -(c-2) S_{1}(y) -(c+1) y \partial_{y} S_{1}(y)
+ 12 
\mbox{ Re } [{\cal F}_{2}^{a}(y) + {\cal F}_{2}^{b}(y) +{\cal F}_{2}^{c}(y)].
\ee
%
%
\section{Connected  piece of the conductivity}

The topologically distinct diagrams resulting from the expansion 
of the connected four-point order-parameter average (\ref{EQ:FOURPT}) to 
$O(u^{2})$ are shown in Fig.\ 4. Self-energy  
corrections are included in these diagrams since we have renormalized the
theory, following dimensional regularization. The algebraic expressions
for each allowed permutation of wavevector and frequency in these diagrams
is inserted in $\sigma^{(4)}$ (\ref{EQ:VERTEX}), 
thereby giving a contribution to the conductivity.  
The $O(u)$ diagram in Fig.\ 4a does not  contribute
to the conductivity since in this case  
%
%
%
the integral (\ref{EQ:VERTEX}) separates into a product of odd
 integrals over ${\bf k}_{1}$ and ${\bf k}_{2}$.  The remaining 
diagrams in Fig.\ 4 are $O(u^{2})$, and produce
%
%
\be
\label{EQ:CONDV}
\sigma^{(4)}(\omega) = - \frac{128 e^2 }{d \Gamma} \kappa^{-\e}
\mbox{ } (u \kappa^{-\e})^{2} \mbox{ } x^{-3 \epsilon} \mbox{ Re } 
[ 4 \tilde{ I}_{b} (y) + \tilde{I}_{c}^{(1)} (y) +
\tilde{I}_{c}^{(2)} (y) ]
\ee
when inserted into Eq.\ (\ref{EQ:VERTEX}). The diagram in 
Fig.\ 4b is responsible for the contribution  
%
%
\be
\tilde{I}_{b} (y) = 
\int_{k_{1} \w_{1} k_{2} \w_{2} k_{3} \w_{3} } \hspace{-.3 in} \vk_{1} \cdot 
\vk_{2} \mbox{ } 
\tilde{G}_{0} (\vk_{1}, \omega_{1}) 
\tilde{C}_{0} (\vk_{1}, \omega_{1} - 2 y)
\tilde{C}_{0} (\vk_{2}, \omega_{2})  
\tilde{C}_{0} (\vk_{2}, \omega_{2} + 2 y) 
\tilde{G}_{0} (\vk_{3}, \omega_{3})
\tilde{C}_{0} (\vk_{1} + \vk_{2} + \vk_{3}, \omega_{1} 
+ \omega_{2} -\omega_{3})
\label{EQ:IB}
\ee
in (\ref{EQ:CONDV}) with $y$ defined in 
(\ref{EQ:Y}) and $\tilde{G}_{0}$ and $\tilde{C}_{0}$ given
by (\ref{EQ:TG0}) and (\ref{EQ:TC0}), respectively.
The diagram in Fig.\ 4c produces the
other two integrals,
%
%
\be
\label{EQ:IC1}
\tilde{I}^{(1)}_{c} (y)  =   \int_{k_{1} \w_{1} k_{2} \w_{2} k_{3} \w_{3} } 
\hspace{-.5 in} \vk_{1} 
\cdot \vk_{2} \mbox{ } \tilde{G}_{0} (\vk_{1}, \omega_{1}) 
\tilde{G}_{0} (\vk_{1},2 y - \omega_{1}) 
\tilde{C}_{0} (\vk_{2},\omega_{2}) 
\tilde{C}_{0} (\vk_{2}, \omega_{2}+ 2 y)
\tilde{C}_{0} (\vk_{3},\omega_{3}) \tilde{C}_{0} (\vk_{1} + \vk_{2} +
 \vk_{3}, \omega_{1} + \omega_{2} + \omega_{3} )
\ee
and
%
%
\be
\tilde{I}^{(2)}_{c} (y)  =  \int_{k_{1} \w_{1} k_{2} \w_{2} k_{3} \w_{3} }
\hspace{-.5 in} \vk_{1} 
\cdot \vk_{2}
\mbox{ } 
\tilde{G}_{0} (\vk_{1}, \omega_{1}) 
\tilde{C}_{0} (\vk_{1},\omega_{1} - 2 y)
\tilde{C}_{0} (\vk_{2}, \omega_{2}) 
\tilde{G}_{0} ( \vk_{2}, - 2 y - \omega_{2} ) 
\tilde{C}_{0} (\vk_{3},\omega_{3})
 \tilde{C}_{0} (\vk_{1} + \vk_{2} +
 \vk_{3}, \omega_{1} + \omega_{2} + \omega_{3} ),
\label{EQ:IC2}
\ee
in (\ref{EQ:CONDV}).
%
%
\begin{figure}
\begin{center}
    \leavevmode
    \epsfysize=2 in
    \epsffile{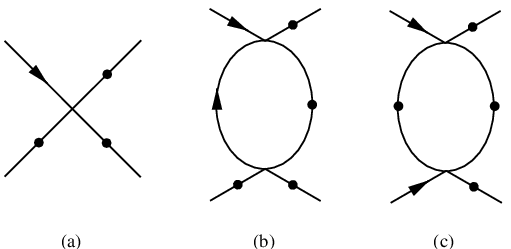}
  \end{center}

{\small FIG. 4. The topologically distinct diagrams in the expansion of 
the four-point order parameter average (\ref{EQ:FOURPT}) to $O(u^{2})$.
The diagrammatic symbols are the same ones used in Fig.\ 3.
Each diagram corresponds to several possible 
wavevector and frequency assignments, 
which are not shown. }
\label{FIG:VERTEX}
\end{figure}

As with the integral
$\tilde{I}_{2} (y)$ (\ref{EQ:TI2}) for the disconnected piece, we 
evaluate the frequency integrals in Eqs.\ (\ref{EQ:IB}-\ref{EQ:IC2})  
with contour integration and use the Feynman formula (\ref{EQ:FEYFORM})
to perform the wavevector integrals. Upon $\e$-expanding the results we 
have
%
%
\bea
\label{EQ:TIB}
\tilde{ I}_{b} (y) & = & - \frac{A_{d}}{96} \left[ 
\left( \frac{1}{4} - \frac{1}{2} \ln \frac{4}{3} \right) \frac{1}{\e^{2}} 
- \left( \frac{3}{4} - \frac{3}{2} \ln \frac{4}{3} \right) 
\frac{1}{\e} \mbox{ Re } [f_{1}(y)] 
+ \frac{0.057}{\e} + {\cal F}_{b} (y) + O(\e) \right], \\
\tilde{ I}_{c}^{(1)} (y) & = & - \frac{A_{d}}{96} \left[ 
\frac{2}{\e^{2}} \ln \frac{4}{3} - \frac{6}{\e} \ln \frac{4}{3} 
\mbox{ Re } [f_{1} (y)]
+ \frac{0.279}{\e} + {\cal F}_{c}^{(1)} (y) + O(\e) \right],  \\
\tilde{ I}_{c}^{(2)} (y) & = &  - \frac{A_{d}}{96} 
\left[ \frac{0.618}{\e} + {\cal F}_{c}^{(2)} (y) + O(\e) \right],
\label{EQ:TIC2}
\eea
where $A_{d}$ and $f_{1}$ were defined in (\ref{EQ:AD}) and  
(\ref{EQ:F1}) respectively. As before, ${\cal F}_{b}$, ${\cal F}_{c}^{(1)}$
and ${\cal F}_{c}^{(2)}$ are $O(\e^{0})$ 
functions of $y$ that do not need to be 
determined. Equations  
(\ref{EQ:TIB}-\ref{EQ:TIC2}) are 
substituted into Eq.\ (\ref{EQ:CONDV}) and the result, expressed in terms of 
a product of $\bar{u}^{2}$ and a series in $\e$, is
%
%
\be
\sigma^{(4)} (\w) =  \frac{e^{2}}{2 \Gamma} \bar{\sigma} \kappa^{-\e}
\mbox{ } 
\bar{u}^{2} \left[ \frac{2}{3 \e^{2}} - \frac{2}{\e} \ln x + \frac{2}{\e} 
S_{1} (y) + \frac{0.086}{\e} - 0.258 \ln x + 3 (\ln x)^{2} - 6 S_{1} (y) \ln x
+ {\cal C}(y) + O(\e) \right],
\label{EQ:S4FIN}
\ee
where
\be
{\cal C}(y) = 0.787 - 2 S_{1}(y) + \frac{2}{3} \mbox{ Re } [ 4 {\cal F}_{b}(y)
+ {\cal F}_{c}^{(1)} (y) + {\cal F}_{c}^{(2)} (y) ].
\ee
In Eq.\ (\ref{EQ:S4FIN}) we have used the relation (\ref{EQ:FS1}) between 
$f_{1}$ and $S_{1}$, Eq.\ (\ref{EQ:S1}).
%
%
\section{Additive Renormalization of the conductivity}
\label{SEC:RENORM}

The real part of the conductivity (\ref{EQ:CONDSPLIT}) is
sum of the disconnected
contributions, Eqs.\ (\ref{EQ:I1FIN}) and (\ref{EQ:I2FIN}), and the 
connected piece, Eq.\  (\ref{EQ:S4FIN}):
%
%
\bea
\sigma'(\w) & = & \frac{e^{2}}{2 \Gamma} \bar{\sigma} \kappa^{-\e}
(1 - 2.6 \bar{u}^{2} ) 
\left\{ \frac{1}{\e} - \frac{2}{3 \e^{2}} c \bar{u}^{2} 
+ \frac{1.4}{\e} \bar{u}^{2} 
-\ln x + \frac{\e + 2 c \bar{u}^{2}}{2} (\ln x)^{2}
- \frac{\e^{2}}{6} (\ln x)^{3} 
 + \Big{[} 1 - (\e + 2 c \bar{u}^{2}) \ln x
\nonumber \right. \\ & & \hspace{-.2 in}
+ \frac{\e^{2}}{2} (\ln x)^{2} \Big{]} S_{1} (y)
+ (\e + 2 c \bar{u}^{2}) (1 - \e \ln x) S_{2} (y)
  + (c+1) (\eta - 2 \bar{u}^{2}) y \partial_{y} S_{1} (y) 
\mbox{ } \ln x + \bar{u}^{2} {\cal F} (y)
+ O(\bar{u}^{2} \e, \e^{3}) \bigg{\}},
\label{EQ:CONDBAR}
\eea
where
\be
{\cal F} (y) = 2.1 S_{1} (y) - 3 c S_{2} (y) + \frac{\e^2}{\bar{u}^{2}}
 S_{3}(y)
+ {\cal D}(y) + {\cal C} (y) - (c+1) y \partial_{y} S_{2} (y)
\ee
is an $O(\e^{0})$ function of $y$.
Even after renormalizing the bare quantities in the theory some poles in
$\e$ remain in Eq.\ (\ref{EQ:CONDBAR}).
 In fact, this problem arises even in the Gaussian theory 
[the $1/ \e$ term in (\ref{EQ:CONDBAR})] and indicates that we must 
be more careful when we set the cutoff $\Lambda$ to infinity.
We should write the conductivity for $d<4$ as  
%
%
\be
\label{EQ:CONDDECOMP}
\sigma' (\w;d,\Lambda) = \sigma' (\w;d,\infty) - A (\w;d,\Lambda),
\ee
with 
%
%
\be
\label{EQ:ADD}
A (\w;d,\Lambda) = \sigma' (\w,d,\infty) - \sigma' (\w;d,\Lambda).
\ee
The 
 $\sigma'(\w;d,\infty)$ term in Eq.\ (\ref{EQ:CONDDECOMP}) is just
Eq.\ (\ref{EQ:CONDBAR}). By subtracting $A$ (\ref{EQ:ADD}) from 
$\sigma'(\w;d,\infty)$ we render the conductivity finite in four 
dimensions, since we recover the theory with finite $\Lambda$. At low
frequencies, near $T_{c}$, we expect to be able to approximate  
$A$ by its value at $T_{c}$ and $\w = 0$: near criticality, the IR
singularities, which appear in $\sigma'(\w;d,\infty)$, are absent in $A$
since only UV physics contributes to the difference in (\ref{EQ:ADD}).
In the minimal subtraction 
scheme
 the poles of $\sigma'(\w=0;d,\infty)$ which contain no singular temperature
dependence are simply subtracted from Eq.\ (\ref{EQ:CONDBAR}). 
This situation is reminiscent of the additive renormalization of the 
specific heat in the static theory \cite{AMIT}. 

Inspection of Eq.\ (\ref{EQ:CONDBAR}) gives
%
%
\be
A = 
\frac{e^{2}}{2 \Gamma} \bar{\sigma} \kappa^{-\e} 
(1 -  2.6 \bar{u}^{2} )  
\left[\frac{1}{\e} - \frac{2}{3 \e^{2}} c \bar{u}^{2}  + 
\frac{1.4}{\e} \bar{u}^{2} + O(\bar{u}^{3}) \right],
\ee
and thus we write the fully renormalized conductivity $\sigma'_{R}(\w) =
\sigma'(\w) - A$ as
%
%
\bea
\sigma'_{R} (\w) & = & \frac{e^{2}}{2 \Gamma} \bar{\sigma} \kappa^{-\e}
(1 - 2.6 \bar{u}^{2}) 
\left\{ -\ln x + \frac{\e + 2 c \bar{u}^{2}}{2} (\ln x)^{2}
- \frac{\e^{2}}{6} (\ln x)^{3}
+ \left[ 1 - (\e + 2 c \bar{u}^{2}) \ln x
+ \frac{\e^{2}}{2} (\ln x)^{2} \right] S_{1} (y)
\nonumber \right. \\ & & \hspace{1 in}
+ (\e + 2 c \bar{u}^{2}) (1 - \e \ln x) S_{2} (y)
  + (c+1) (\eta - 2 \bar{u}^{2}) y \partial_{y} S_{1} (y) 
\mbox{ } \ln x  
+ \bar{u}^{2} {\cal F}(y)
+ O(\bar{u}^{2} \e, \e^{3}) \Bigg{\}}.
\label{EQ:RENCOND1}
\eea
\vspace{-1.7cm}\noindent\hspace{9.2cm}\makebox[8.6cm]{\hrulefill}\vspace{1.3cm}
\begin{multicols}{2} 
\hspace{-.2 in} 
Now we have a theory that is UV convergent as $\e \rightarrow 0$, 
but has IR divergences as $T \rightarrow T_{c}$ and 
$x \rightarrow 0$. Near four dimensions, the coupling constant $\bar{u}$ 
flows in the IR to its XY-model fixed-point value $\bar{u}^{*}$ 
(\ref{EQ:BFIXED}), with $\eta= 2 (\bar{u}^{*})^{2}$ (\ref{EQ:B-ETA}) 
and, after re-summing, the series in $\e$ takes the form 
%
%
\bea
\label{EQ:SIGALMOST}
\sigma'_{R} (\w) & = &   \frac{e^{2}}{2 \Gamma} \bar{\sigma} \kappa^{-\e} 
\Big{[} \frac{x^{-p} -1}{p} + x^{-p} S_{1}(y) 
 \nonumber \\ & & \hspace{.5 in} 
+ p x^{-p} S_{2} (y)
+  p^{2} {\cal F} (y) + O(p^{3}) \Big{]},
\eea
where $p$ is $O(\e)$ and is defined as 
\bea
p & = & \e + c \eta \nonumber \\
  & = & 2 - d + z + O(\e^{3}) \nonumber \\
  & = & \frac{2}{z} (2 - d + z) + O(\e^{3}).  \nonumber
\eea
The $(1 - 2.6 \bar{u}^{2})$ factor in Eq.\
(\ref{EQ:SIGALMOST}) has been absorbed by changing 
the normalization of $\sigma'_{R} (\w=0)$.
As $T \rightarrow T_{c}$ and  $x \rightarrow 0$ terms
proportional $x^{-p}$ in Eq.\ (\ref{EQ:SIGALMOST}) dominate 
the conductivity \cite{CAVEAT}. From Eq.\ (\ref{EQ:XTEMP}) we have
\be
x^{-p} = (\xi \kappa)^{p} [ 1 + O(\e^{3})],
\ee
and as $T \rightarrow T_{c}$ we write
\be
\sigma'_{R} (\w) =  \frac{e^{2}}{2 \Gamma} \bar{\sigma} \kappa^{-\e}
\frac{ (\xi \kappa)^{p}}{p} [ 1 + p S_{1}(y) + p^{2} S_{2}(y) + O(p^{3})].
\label{EQ:CONDLAST}
\ee
In Eq.\ (\ref{EQ:CONDLAST}), the series in $p$ coincides,
to $O(p^{2})$, with the $\e$-expansion 
 for the Gaussian scaling 
form $S'_{G}$, Eq.\ (\ref{EQ:EEXPG}). Thus, by re-summing the series in
Eq.\ (\ref{EQ:CONDLAST}) we obtain, correct to $O(\e^{2})$, 
the Gaussian scaling form $S'_{G}$, Eq.\ (\ref{EQ:CGR}),
 now  as a function of the 
critical scaled frequency $y$ (\ref{EQ:Y}) and
 with occurrences of $\e$ replaced by $p$.

The final result for complex a.c.\ conductivity 
in the critical regime is then (dropping the $R$ suffix)
\vspace{.1 in} \\
%
%
\be
\label{EQ:FINALRESULT}
\sigma (\w) = \frac{e^{2}}{2 \Gamma} \bar{\sigma} \kappa^{-\e}
\mbox{ }
\frac{(\xi \kappa)^{2-d + z}}{(2-d + z)} 
\mbox{ } 
[S(y) + O(\e^{3})],
\ee
with the scaled frequency $y$ given by Eq.\ (\ref{EQ:Y})
and the universal complex scaling function $S(y)$ given by 
Eq.\ (\ref{EQ:SINTRO}).
%
%
\section{Comparison with experiment}
\label{SEC:COMPARE}

It is instructive to compare the universal function $S(y)$, Eq.\ 
(\ref{EQ:SINTRO}), for the critical theory (extrapolated to $d=3$), 
with both the 
prediction of the Gaussian theory,
Eq.\ (\ref{EQ:CGI}), and the experimental results of Ref.  
\cite{BOOTH96}. Strictly speaking, it is inconsistent to compare 
scaled data from different theories and experiments if the axes have
been scaled using different exponents. However, for the sake of comparison,
we take the viewpoint
that the theory and  experiment each determine a particular 
universal functional dependence $S(y)$ and ignore exactly 
how $S(y)$ and $y$ are achieved.

In this spirit, the magnitude of $S(y)$ as a function of $y$
is plotted  on a $\log$-$\log$ scale 
in Fig.\ 1 for the critical
and Gaussian theories. Since $z \stackrel{\textstyle >}{_{\sim}} 2$
in the critical theory,
the power-law behaviour at large $y$ [a consequence of (\ref{EQ:CONTC})]
for the critical theory lies only slightly below the Gaussian theory. 
%
%
\begin{figure}
\begin{center}
    \leavevmode
    \epsfxsize=3.25 in
    \epsffile{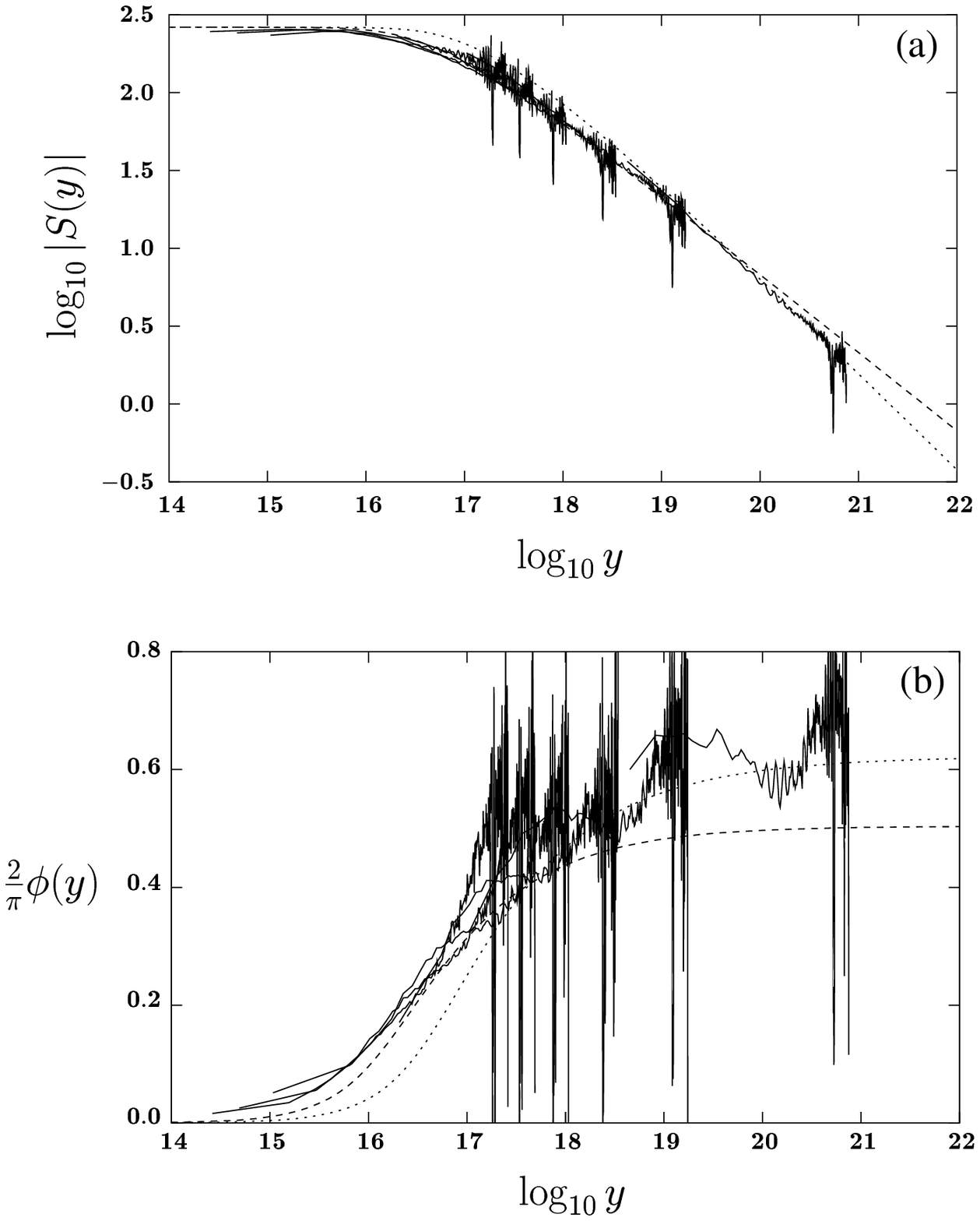}
  \end{center}

{\small FIG. 5. Comparison between the scaled a.c.\ conductivity data from 
Booth {\em et al.},
Ref. \protect\cite{BOOTH96}, on YBCO and the relaxational 3D XY critical 
theory. (a) The scaling function $S(y)$, Eq.\ (\ref{EQ:SINTRO}), using
the relaxational 3D XY value $z=2.015$ (dashed curve) and using the
experimental value $z=2.65$ (dotted curve) are compared with the 
experimental results (solid curves). The magnitude of $S(y)$ is 
plotted against $y$ on a $\log-\log$ scale. The theory is fit to the
experiment using horizontal and vertical offsets (the horizontal
offset depends on the value of $z$ used). (b) The normalized
phase, $ 2\phi(y)/ \pi$, of the conductivity is plotted against 
$\log_{10} y$ for the relaxational 3D XY critical theory with $ z=2.015$
(dashed curve), the theory using the experimental value $z=2.65$ (dotted curve)
and experiment (solid curves). The horizontal offsets are the same as in (a).}
\label{FIG:EXPERCOMP}
\end{figure}
 
In Fig.\ 5, the critical theory is compared with measurements
of the microwave conductivity of a thin 
film sample of YBCO in the range 45 MHz - 45 GHz
near $T_{c}$ \cite{BOOTH96}. 
In this experiment, the exponent $z=2.65 \pm 0.3$ and the 
transition temperature $T_{c} = 89.1 \pm 0.1$ K, were 
determined from the power-law behaviour (\ref{EQ:CONTC}) expected
at $T_{c}$. The best scaling collapse of the data determined 
the value $\nu = 1.0 \pm 0.2$ for the static exponent.  In Fig.\ 
5a the magnitude of $S(y)$ is again plotted as a 
function of $y$ on a log-log scale. 
The Gaussian theory is not plotted since it lies 
so close to the critical theory. Since  $\Gamma$, 
$\kappa$ and the prefactor of $\xi$, which appear in both the scaled 
frequency $y$ (\ref{EQ:Y}) and the prefactor to the conductivity 
(\ref{EQ:FINALRESULT}), are parameters in 
the TDGL theory, there is freedom 
to choose the horizontal and vertical positioning of the theory so as to give 
the best fit to the data. As with the Gaussian theory, the critical theory 
fits the experimental scaling curve well over almost four decades in 
scaled frequency $y$, but deviates from the experimental data taken 
nearest to $T_{c}$. 

The dynamic exponent for the relaxational 3D XY-model is known 
to have the value  $z \approx 2.015$. Nevertheless,
it is instructive to consider the $z$ 
appearing in $S(y)$, Eq.\ (\ref{EQ:SINTRO}), as an adjustable parameter. 
By choosing the experimental value $z=2.65$ and adjusting the horizontal
offset of the theory in Fig.\ 5, a better fit to the 
experimental data closest to $T_{c}$ is achieved---at the expense of worse 
agreement with the rest of the data. This comparison emphasizes that 
the experimental value $z=2.65$ seems to originate in the data set 
taken closest to $T_{c}$.

The phase $\phi(y)$, Eq.\ (\ref{EQ:PHASEDEFN}), of the conductivity is 
plotted against $\log_{10} y$ in Fig.\ 5b
for the critical theory ($z=2.015$), ``pseudo''-theory ($z=2.65$) and
the experiment \cite{BOOTH96}. As with the Gaussian theory,
the critical theory predicts a smaller phase near $T_{c}$ than seen 
experimentally.
The ``pseudo''-theory is in better agreement with experiment near $T_{c}$ 
than the critical theory, but again does a poorer job fitting the rest of the 
curve.
%
%

\section{Conclusions}
\label{SEC:CONCLUSIONS}

We have examined a theory for the a.c.\ conductivity 
of a superconductor that includes the strong, interacting 
order-parameter fluctuations
expected near criticality. The FFH scaling hypothesis, Eq.\
 (\ref{EQ:SCALCON}), 
is shown to hold at $O(\e^{2})$ in the $\e$-expansion for relaxational  
3D XY-model critical dynamics.
The universal 
scaling function $S(y)$ appearing in Eq.\ (\ref{EQ:SCALCON}) is explicitly 
calculated to $O(\e^{2})$ for this dynamics, with the result 
given in Eq.\ (\ref{EQ:SINTRO}). The frequency 
and phase behaviour expected at $T_{c}$, 
Eqs.\ (\ref{EQ:CONTC}) and (\ref{EQ:PHASE}) respectively,  
is demonstrated. The critical scaling function $S(y)$
generalizes the Gaussian result, Eq.\ (\ref{EQ:CGI}), and reduces to it when 
$z=2$. These results are quite general and hold, in the critical 
regime, for any bulk superconductor
described by a complex order-parameter with relaxational dynamics.

Since $z \approx 2$ for this dynamics, the scaling  
curve $S(y)$ is, for practical purposes, indistinguishable 
from the prediction of the Gaussian theory (see Fig.\ 1). 
Therefore, in a 
measurement of the a.c.\ conductivity, the only indication
of a crossover from the Gaussian to critical fluctuation regime would 
be a crossover in the static exponent $\nu$. 
This may explain why the Gaussian theory fits the 
experimental data of Booth {\em et al.}, Ref. \cite{BOOTH96},
 so well over much of 
the curve in Fig.\ 5, 
even though the experiment is supposedly accessing 
the critical regime.

The inclusion of critical order-parameter fluctuations in the
framework of  relaxational dynamics does not seem sufficient 
to explain the deviation between the 
Gaussian scaling form and experiment \cite{BOOTH96} observed near $T_{c}$
(see Fig.\ 5). As highlighted by the fit of the
``pseudo''-theory in Fig.\ 5, this deviation
is connected to the large value $z=2.65$ obtained in the experiment, which
cannot explained within any present theory \cite{POSSIBILITIES}.
It is possible that this discrepancy 
may be due to the strong influence
that uncertainties in the experimental determination of $T_{c}$ have
on the scaling of the data closest to $T_{c}$. More 
a.c.\ conductivity measurements with higher temperature resolution near 
$T_{c}$ may resolve this issue, allow a more accurate determination of $z$,
and provide a check on the scaling collapse for large $y$.
It is also possible that the films studied contain strong disorder, 
which could affect the scaling near $T_{c}$. 

In this paper we have identified and dealt with the technical challenges
involved in the organization and renormalization of the theory for the a.c.\
conductivity in the critical region. This work serves as a basis for 
examining more complicated models, such as model F of Hohenberg and Halperin
\cite{HH} involving reversible couplings to a conserved energy-mass 
density field, as in superfluid $^{4}$He. In three dimensions 
$z=3/2$ for model F \cite{DEDOMINICIS78,HALPERIN76} 
which, although not observed in the a.c.\
conductivity data \cite{BOOTH96}, is seen in some d.c.\ conductivity 
experiments \cite{POMAR93,HOLM95,KIM97} and simulations
\cite{MONTECARLO}.
Another extension of the present theory is to consider a
non-zero magnetic field, with the aim of examining the crossover 
from the zero-field critical scaling of the 3D XY-model to the 
lowest-Landau-level scaling which obtains in high fields 
\cite{MOLONI97a,ULLAH91}. 
%
%
\acknowledgements

The authors would like to thank Gene Mazenko and 
Andrei Varlamov for useful comments, and 
Steve Anlage for providing 
his experimental data. This work was supported by NSERC of Canada and 
NSF grant DMR 96-28926.
\end{multicols}
%
%
\appendix
\section{}
\label{APPENDIX}

To illustrate the calculation of wavevector integrals, 
we use this appendix to provide the
details of the $\e$-expansion of $\tilde{I}_{2}^{a} (y)$,
Eq.\ (\ref{EQ:I2a}), which is reproduced, in the notation of
Sec.\ \ref{SEC:DIS2}, as
\be
\label{EQ:Q1}
\tilde{I}_{2}^{a} (y) =  \int_{0}^{1} d v \mbox{ } (1 - v)
\int_{k_{1} k_{2} k_{3}} k_{1}^{2} \mbox{ } \left[ \frac{2}{a_{1}^{3}}
- \frac{1}{(a_{1} + i y v)^{3}} 
\right] \frac{1}{a_{2} a_{3} (a_{2} + a_{3} 
+ a_{4}) (a_{5} + 2 i y)}. 
\ee
We parameterize the wavevector factors in the denominator of Eq.\
(\ref{EQ:Q1}) in pairs using the Feynman parameterization 
(\ref{EQ:FEYFORM}), beginning with factors on the right containing  $\vk_{3}$.
The denominator of the $\vk_{3}$ integral is thereby transformed into a 
quadratic form in  $\vk_{3}$  and the integral is solved. The 
process is repeated for the remaining two  wavevector integrals,
producing
\be
\tilde{I}_{2}^{a} (y) = \frac{A_{d}}{3 \e} \int_{0}^{1} d v 
\mbox{ } (1-v) \mbox{ }  [ 2 J (v=0,y) - J (v,y)]
\label{EQ:I2-J}
\ee
with 
\be
\label{EQ:JV}
J (v,y) = \int_{0}^{1} d u_{1} d u_{2} d u_{3} d u_{4} \mbox{ } u_{2} 
(1 + u_{2})^{\epsilon-1} u_{3}^{\epsilon/2}
(1-u_{4})^{2} u_{4}^{\epsilon -1}  
\frac{\tilde{g}_{0}^{-3 \epsilon/2}}{g_{2}^{2 - \epsilon/2} 
\tilde{g}_{1}^{3 - \epsilon/2}},
\ee
where we have defined
\bea
\tilde{g}_{0} & = & (1- u_{4}) (1 + i y v) + u_{4} g_{0}, \\
\tilde{g}_{1} & = & 1 + u_{4} (g_{1} -1 ), \\
g_{0} & = & 1 - u_{3} + u_{3} (1 + u_{2}) \{1 + u_{2} [2 + u_{1}
(1 + 2 i y)] \}, \\
g_{1} & = & \frac{u_{2} u_{3}}{g_{2}} \{ g_{2} [1 + u_{1} (1 + u_{2})] - 
u_{2} u_{3} \}, \\
g_{2} & = & 1 - u_{3} + u_{2} u_{3} (2 + u_{2}). 
\eea
In the $\e$-expansion $J$ in Eq.\ (\ref{EQ:JV}) 
is $O(\epsilon^{-1})$ at leading order. 
 The singularity in $\e$ is isolated by writing $J$ as
\be
\label{EQ:JDECOMP}
J(v,y) = J_{a} (v,y) + J_{b} (v,y)
\ee
where
\bea
\label{EQ:JA}
J_{a} (v,y) & = & 
(1 + i y v)^{-3 \e/2} 
\int_{0}^{1} d u_{2} d u_{3} d u_{4} \mbox{ } u_{2} 
(1 + u_{2})^{\epsilon-1} \frac{u_{3}^{\epsilon/2}}{g_{2}^{2 - \epsilon/2}}
 u_{4}^{\epsilon -1}, 
\\
J_{b} (v,y) & = & \int_{0}^{1} d u_{1} d u_{2} d u_{3} d u_{4} \mbox{ } u_{2} 
(1 + u_{2})^{\epsilon-1} \frac{u_{3}^{\epsilon/2}}{g_{2}^{2 - \epsilon/2}}
 u_{4}^{\epsilon - 1} 
\left[ (1-u_{4})^{2} 
\frac{\tilde{g}_{0}^{-3 \epsilon/2}}{\tilde{g}_{1}^{3 - \epsilon/2}} - 
(1+ i y v)^{-3 \e /2} 
\label{EQ:JB}
\right].
\eea
In the $\epsilon$-expansion, Eq.\ (\ref{EQ:JA}) becomes
\bea
J_{a}(v,y) & = & \frac{1}{\epsilon} \int_{0}^{1} d u_{2} d u_{3}
\mbox{ } \frac{u_{2}}{g_{2}^{2} (1 + u_{2})}
- \frac{3}{2} \mbox{ } \ln (1 + i y v) 
\mbox{ } \int_{0}^{1} d u_{2} d u_{3}
\mbox{ } \frac{u_{2}}{g_{2}^{2} (1 + u_{2})}
\nonumber \\ & & \hspace{1 in}
+ \int_{0}^{1}  d u_{2} d u_{3}
\mbox{ } \frac{u_{2}}{g_{2}^{2} (1 + u_{2})} \left[ \ln (1 + u_{2}) 
+ \frac{1}{2} \ln u_{3} + \frac{1}{2} \ln g_{2} \right]
+ \e {\cal F}_{a} (v,y) + O(\epsilon^{2}),
\label{EQ:JAE}
\eea
where ${\cal F}_{a} (v,y)$ is a function of $v$ and $y$.
The integrals in Eq.\ (\ref{EQ:JAE}) are evaluated to produce
\be
J_{a}(v,y) = \frac{1}{\epsilon} \ln \frac{4}{3} 
- \frac{3}{2} \ln \frac{4}{3} \times \ln (1 + i y v) 
 - 0.087 + \e {\cal F}_{a} (v,y) 
+ O(\epsilon^{2}).
\label{EQ:JAFIN}
\ee
The non-singular integral $J_{b}$, Eq.\ (\ref{EQ:JB}), 
has the expansion
\bea
J_{b} (v,y) & = & \int_{0}^{1} d u_{1} d u_{2} d u_{3} d u_{4} \mbox{ } 
\frac{u_{2}}{g_{2}^{2}(1 + u_{2})} \frac{1}{u_{4}} 
\left[ \frac{(1-u_{4})^{2}}{\tilde{g}_{1}^{3}} - 1 \right]
+ \epsilon {\cal F}_{b} (v,y) + O(\e^{2})
\nonumber \\ & & 
\nonumber \\ 
& = & 0.103 + \epsilon {\cal F}_{b} (v,y) + O(\e^{2}),
\label{EQ:JBFIN}
\eea
where ${\cal F}_{b} (v,y)$ is a function of $v$ and $y$.
By combining the Eqs.\ (\ref{EQ:JAFIN}) and
(\ref{EQ:JBFIN}) in Eq.\ (\ref{EQ:JDECOMP}) we may 
use this result for $J$ in $\tilde{I}_{2}^{a}$, Eq.\ (\ref{EQ:I2-J}), to 
obtain
the result quoted in Eq.\ (\ref{EQ:I2aE}), with
\be
{\cal F}_{2}^{a} (y) = \frac{1}{3} \int_{0}^{1} d v (1-v) \{2 
[{\cal F}_{a}(0,y)+ 
{\cal F}_{b} (0,y)] - 
[{\cal F}_{a}(v,y)+ {\cal F}_{b}(v,y)] \}.
\ee
\begin{multicols}{2}
%
%

\end{multicols}

\begin{references}
\bibitem{SUPERFLUID}  Another dynamical model proposed to describe the 
superconducting transition is  model F in the Hohenberg and Halperin 
classification \cite{HH}, where the order-parameter is coupled to a 
reversible, conserved energy-mass density field, as in a superfluid. However, 
the second-sound mode produced by this coupling is expected to be gapped 
in the presence of the Coulomb interaction, suggesting that the
critical dynamics of the superconductor
may be appropriately described using the 
purely relaxational model A \cite{FFH}.
\bibitem{BETADEFN} Our  expressions for the
renormalization constants differ slightly 
from those in \cite{DEDOMINICIS78,AMIT} 
since the coupling $\bar{u}$ defined in Eq.\ (\ref{EQ:DIMCOUP}) 
 involves different numerical 
factors than the definition in \cite{DEDOMINICIS78,AMIT}.
\bibitem{CAVEAT} In Eq.\ (\ref{EQ:SIGALMOST}) it is possible that 
terms of the form $p^{2} x^{-p} \mbox{ }  {\cal F} (y)$
 also contribute to the 
conductivity near $T_{c}$. To check this would, however, require a 
calculation correct to $O(p^{3})$.
\end{references}
\end{document}